\documentclass[reprint,aps,prb,floatfix,twocolumn,superscriptaddress,showpacs,amsmath,amssymb,showpacs,]{revtex4-1}
\usepackage{graphicx}%
\usepackage{bm}
\usepackage[USenglish]{babel} 
\usepackage[utf8]{inputenc}

\usepackage{xcolor}
\usepackage{amsmath}
\usepackage{epsfig}
\usepackage{dcolumn}
\usepackage{amssymb}

\begin{document}

\author{Philipp Werner }
\affiliation{Department of Physics, University of Fribourg, 1700 Fribourg, Switzerland}
\author{Francesco Petocchi} 
\affiliation{Department of Physics, University of Fribourg, 1700 Fribourg, Switzerland}
\author{Martin Eckstein}
\affiliation{Department of Physics, University of Erlangen-Nuremberg, 91058 Erlangen, Germany}

\title{Time-resolved photoemission and RIXS study of a site-selective Mott insulator}

\begin{abstract}
Inspired by the physics of rare earth nickelates, we study the photoemission (PES) and resonant inelastic X-ray scattering (RIXS) spectra of a correlated electron system with two types of insulating sublattices.  Sublattice A is characterized by a hybridization gap and a low-spin state, while sublattice B features a Mott gap and a local magnetic moment. We show how the coupling of these two qualitatively different insulating states affects the dynamics of photo-induced charge carriers and how the nonequilibrium states manifest themselves in the PES and RIXS signals. In particular, we find that charge carriers created on the B sublattice migrate to the A sublattice, where they contribute to the creation of in-gap states in the PES signal, and to characteristic peaks in the nonequilibrium RIXS spectrum. While the contributions from the two sublattices cannot be easily distinguished in the local photoemission spectrum, the weights of the RIXS signals in the two-dimensional $\omega_\text{in}$-$\omega_\text{out}$ space provide information on the local state evolution on both sublattices. 
\end{abstract}

\date{\today}

\maketitle

\section{Introduction}

The rare earth nickelates ReNiO$_3$ with Re = Sm, Eu, Y or Lu exhibit a low temperature paramagnetic insulating phase with bond disproportionation,\cite{Medarde1997,Alonso1999} which has been the subject of numerous theoretical investigations.\cite{Mizokawa2000,Park2012,Lau2013,Johnston2014} Park {\it et al.} \cite{Park2012} proposed the picture of a site-selective Mott phase, where the Ni $d$ electrons on one sublattice are in a paramagnetic Mott insulating state and feature a large local moment, whereas the $d$ electrons on the other sublattice form singlet states with the holes on the surrounding oxygens. Since the bond disproportionation and site selectivity is absent in the metallic high-temperature equilibrium phase, an interesting question concerns the mechanism which leads to this symmetry-broken low-temperature state without magnetic order or charge order on the Ni sites. The authors of Refs.~\onlinecite{Mazin2007,Subedi2015} proposed that an electronic instability can trigger this phase transition if $U-3J$ (with $U$ the Hubbard interaction for the Ni $e_g$ states and $J$ the Hund coupling) is small compared to the energy difference between the inequivalent nickel sites, or even negative.

Here, we will focus on the bond-disproportionated phase and ask what kind of nonequilibrium states can be induced in such a system with coexisting high-spin and low-spin configurations. In particular, we investigate purely electronic excitation processes (photo-doping),  assuming that the lattice remains frozen. We consider a minimal model of a site-selective insulator and use the nonequilibrium extension of dynamical mean field theory (DMFT)\cite{Georges1996,Aoki2014} to investigate the time-resolved photoemission spectroscopy (PES) and resonant inelastic X-ray scattering (RIXS) signals of the photo-doped insulating phase. We will show that the coexistence of Kondo insulating and Mott insulating states on two sublattices results in a nontrivial recombination dynamics of the photo-carriers, which can be tracked by time-resolved PES and RIXS measurements. 

The paper is organized as follows. Section~\ref{sec:model} introduces the model, Sec.~\ref{sec:results} presents the equilibrium and nonequilibrium PES and RIXS spectra, while Sec.~\ref{sec:conclusions} contains a short discussion and conclusions. 
 
\section{Model and Method} 
\label{sec:model}
 
\begin{figure}[t]
\begin{center}
\includegraphics[angle=0, width=0.8\columnwidth]{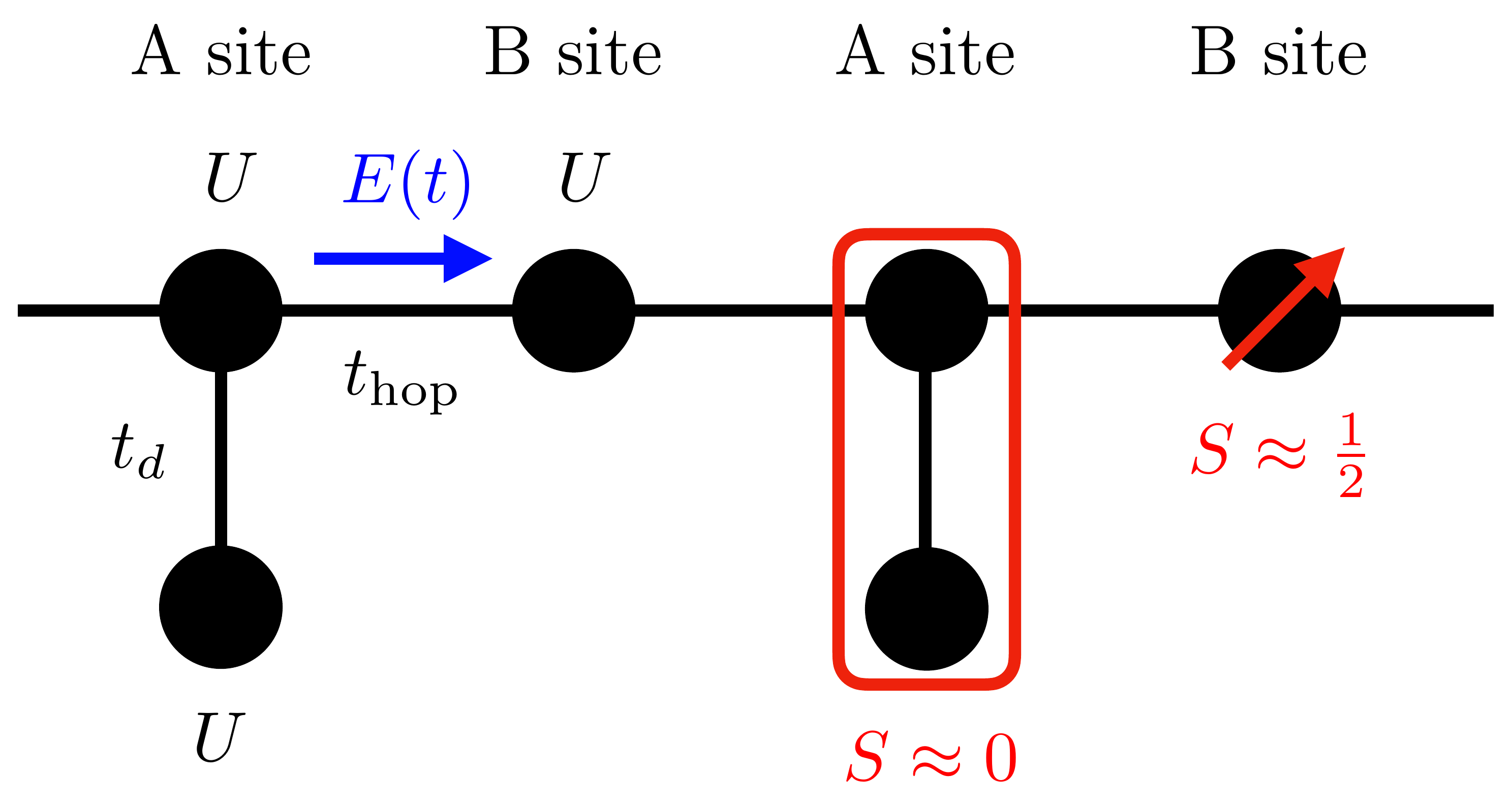} %
\caption{Illustration of the model. The sites on sublattice A consist of dimers with hopping strength $t_d$ and the sites on sublattice B consist of single orbitals. All orbitals are half-filled with an interaction $U$ and $t_\text{hop}$ is the hopping between A and B sublattice sites. For large enough $t_d$ and $U$, the dimers will be predominantly in a spin 0 singlet state, while the B sublattice sites will feature a spin-$\tfrac12$ moment. Nonequilibrium states are induced by an electric field pulse $E(t)$ directed along the $t_\text{hop}$ bonds.  
}
\label{fig_model}
\end{center}
\end{figure}    

The model considered in this study is illustrated in Fig.~\ref{fig_model}. It features two sublattices A and B, with sublattice A sites consisting of dimers with intra-dimer hopping $t_d$, whereas the sites on sublattice B host single orbitals. All the orbitals are on average half-filled and interacting with an intra-orbital interaction $U$, i.e., the local Hamiltonians are $H_\text{A,loc}=U(n_{\text{A},\uparrow}n_{\text{A},\downarrow}+n_{\text{L},\uparrow}n_{\text{L},\downarrow})-t_d\sum_\sigma(d^\dagger_{\text{A},\sigma}d_{\text{L},\sigma} +\text{h.c.})-(\mu+\frac{U}{2})(n_A+n_L)$ and $H_\text{B,loc}=U n_{\text{B},\uparrow}n_{\text{B},\downarrow}-(\mu+\frac{U}{2})n_B$, with $d^\dagger_{\alpha,\sigma}$ denoting the creation operator for an electron with spin $\sigma$ in orbital $\alpha$, $n_{\alpha,\sigma}=d^\dagger_{\alpha,\sigma}d_{\alpha,\sigma}$ the spin density in orbital $\alpha$, $n_\alpha=n_{\alpha,\uparrow}+n_{\alpha,\downarrow}$ the total density in orbital $\alpha$, and $\mu$ ($=0$) the chemical potential. The hopping $t_\text{hop}$ between the sublattice sites only connects to one of the dimer orbitals (index A), so that the other orbital (index L) acts, for large enough $U$, like the spin of a Kondo-lattice model. In the following discussion we will refer to the sites connected by $t_\text{hop}$ as the ``backbone" and to the dangling sites connected to the backbone by $t_d$ as the ``ligands."\cite{footnote_L} In the context of the rare earth nickelates, the A sublattice backbone sites correspond to the Ni $d$ atoms with the short Ni-O bond length (low spin), and the B sublattice sites to those with the long Ni-O bond length (high spin). Indeed, for large enough $t_d$ and $U$, we expect that the dominant local state on sublattice A is a singlet state, while on sublattice B it will be a spin-$\tfrac12$ state, see the right unit cell in Fig.~\ref{fig_model}. In a more realistic model for nickelates, the high-spin state would be $S=1$ and we should consider two $e_g$ orbitals per backbone site, but we are interested here in a minimal toy model which allows to capture the interplay between high-spin and low-spin subsystems in equilibrium and nonequilibrium states. Our model is also related to the physics of the layered transition-metal dichalcogenide 1$T$-TaS$_2$, which depending on the stacking arrangement contains combinations of band-insulating bi-layers and Mott insulating mono-layers.\cite{Ritschel2018,Lee2019,Petocchi2022}

\begin{figure}[t]
\begin{center}
\includegraphics[angle=-90, width=0.9\columnwidth]{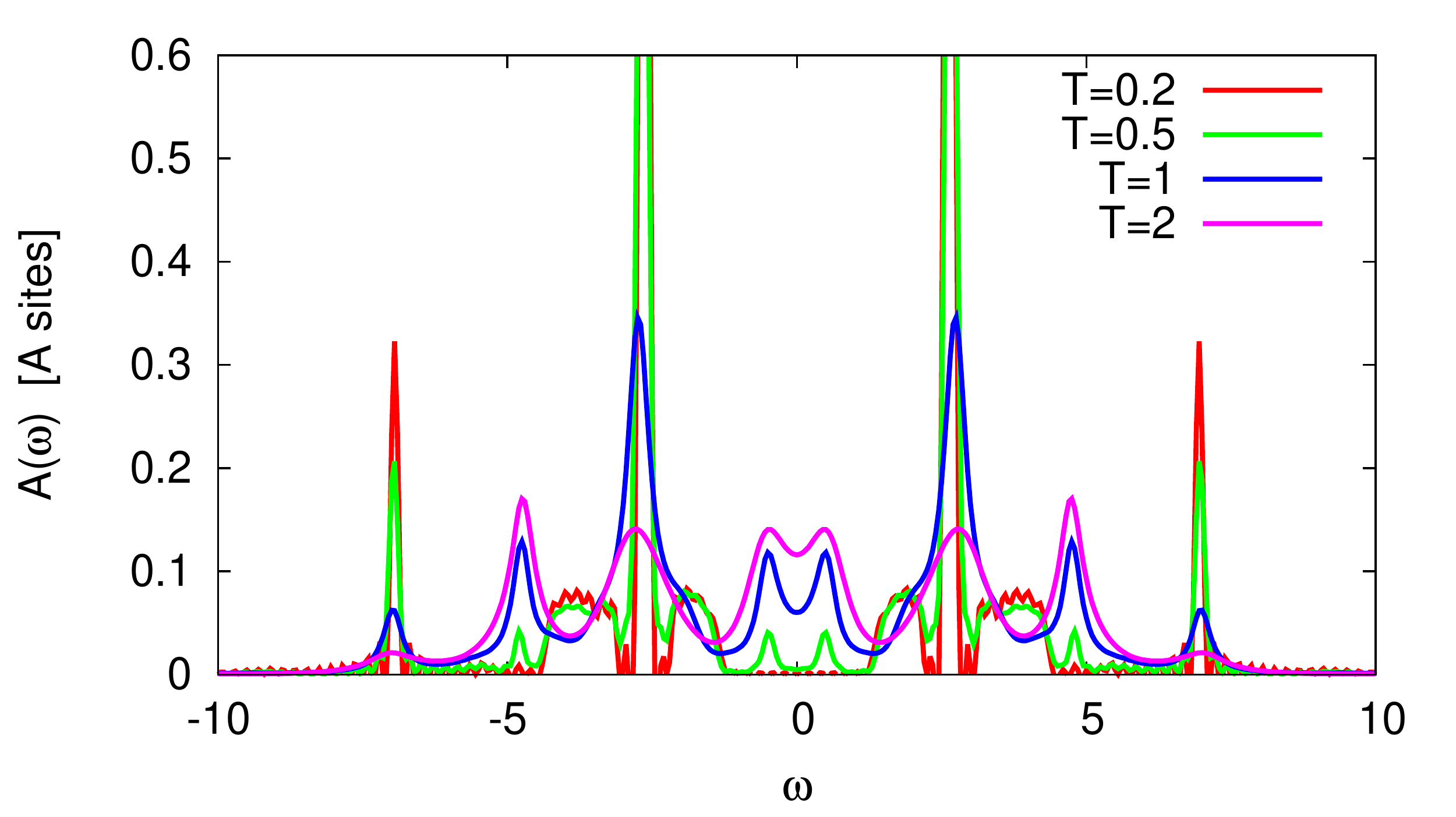} 
\includegraphics[angle=-90, width=0.9\columnwidth]{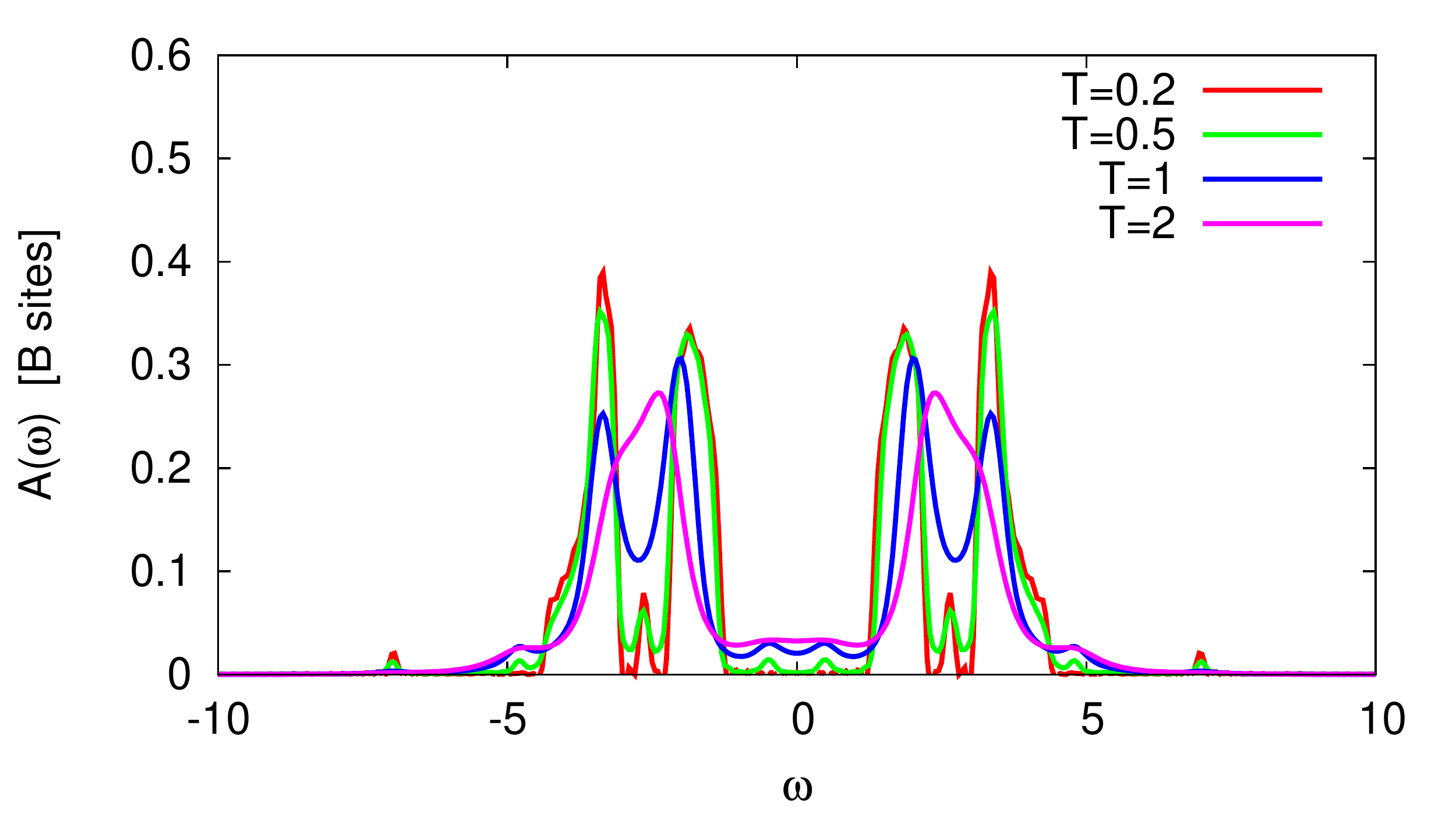} 
\caption{Equilibrium spectral functions for the backbone orbitals on the A sublattice (top panel) and B sublattice (bottom panel) at the indicated temperatures $T$.  The parameters are $U=5$ and $t_d=2$. 
}
\label{fig_eq}
\end{center}
\end{figure}    

\begin{figure}[t]
\begin{center}
\includegraphics[angle=0, width=0.9\columnwidth]{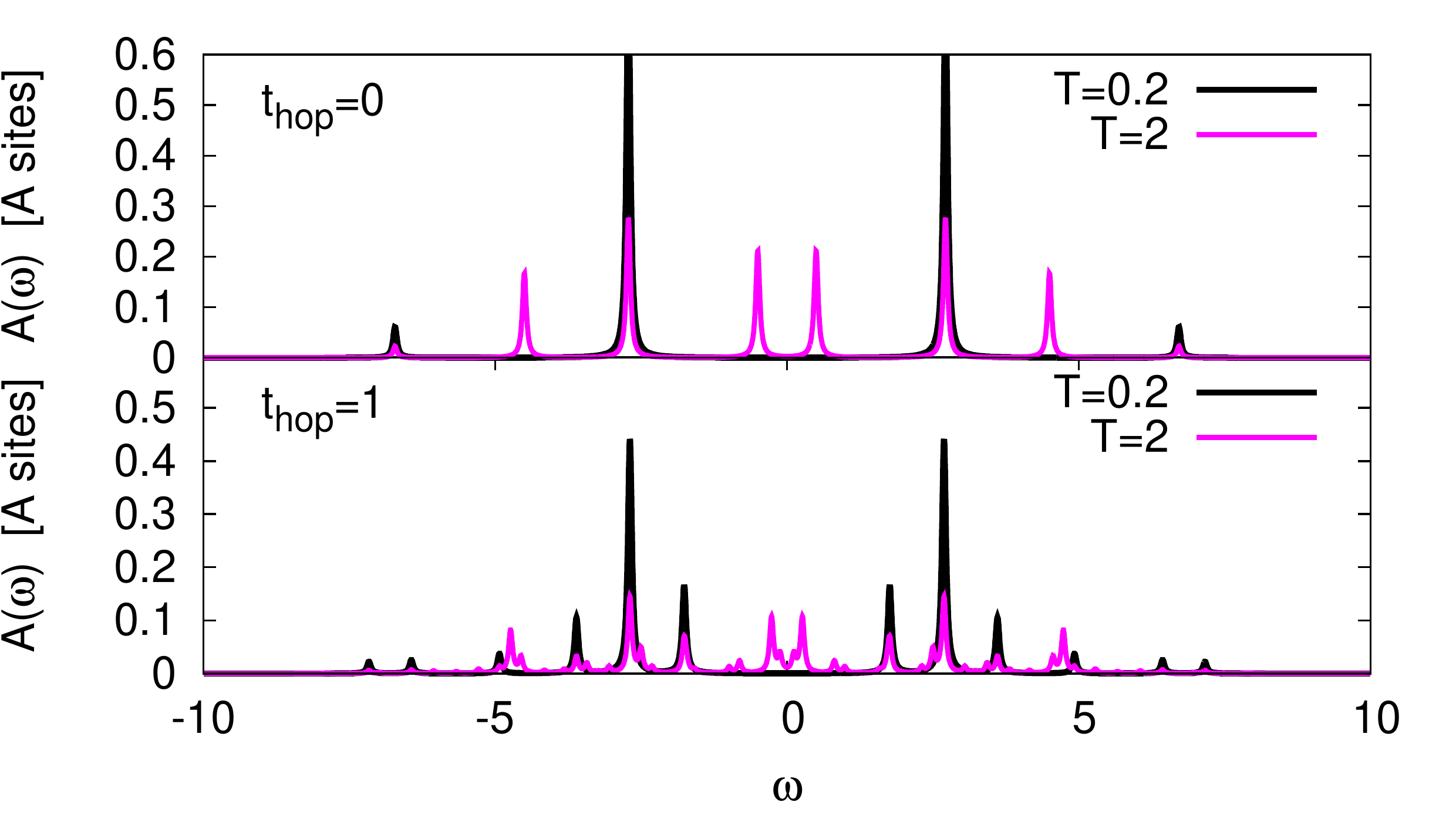}\\ 
\includegraphics[angle=0, width=0.9\columnwidth]{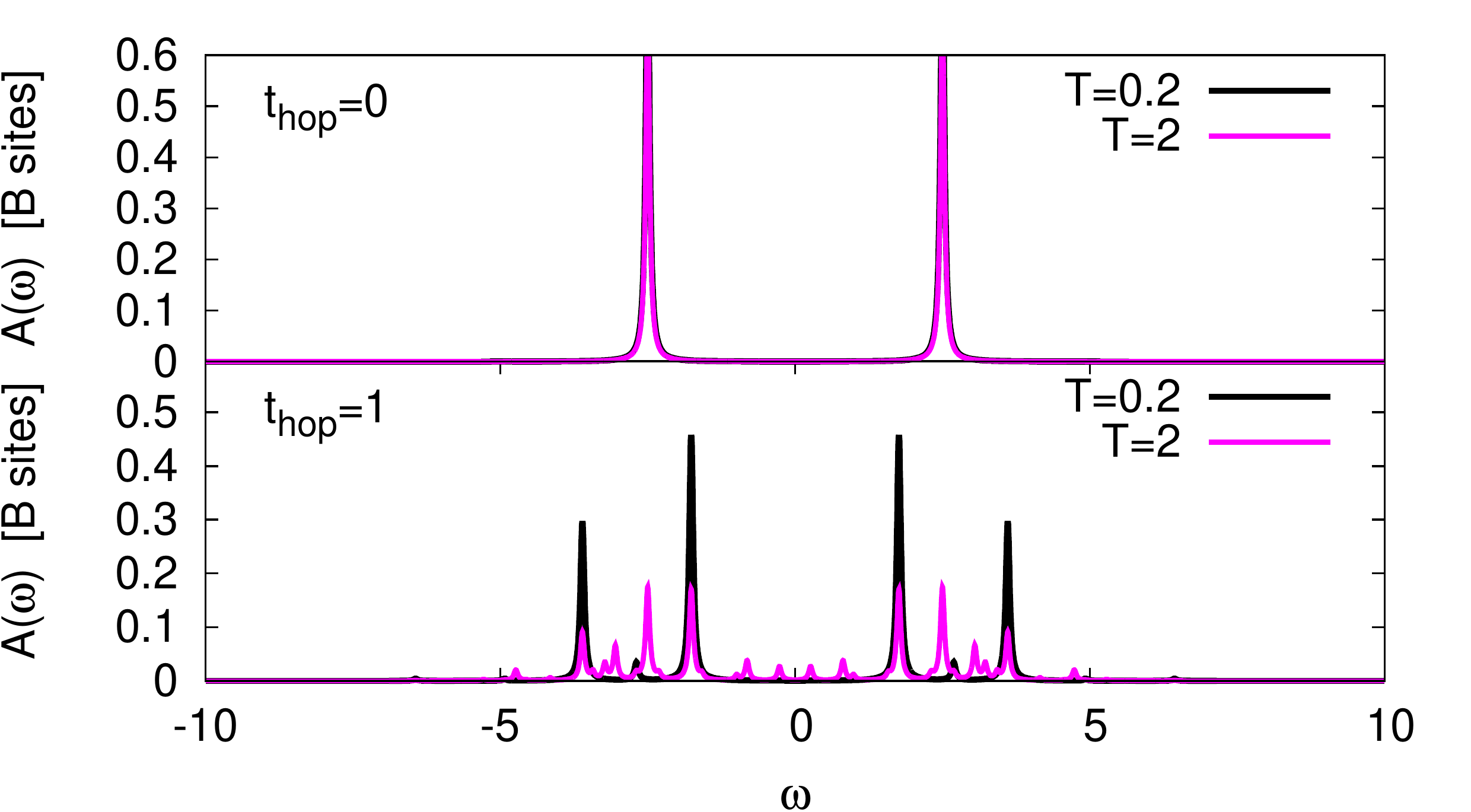} 
\caption{Exact diagonalization spectra for the backbone orbitals on the A sublattice (top panel) and B sublattice (bottom panel) at the indicated temperatures $T$.  The parameters are $U=5$ and $t_d=2$. The upper subpanel plots the atomic solution ($t_\text{hop}=0$) and the lower subpanel the solution of a three orbital model consisting of the two atomic systems connected by a hopping term $t_\text{hop}=1$. 
}
\label{fig_eq_ED}
\end{center}
\end{figure}    

While the illustration in Fig.~\ref{fig_model} is for a one-dimensional system, in the simulations, we will treat an infinitely connected Bethe lattice, with an appropriately renormalized $t_\text{hop}$.\cite{Metzner1989}  Within DMFT, the problem then maps to two coupled impurity models, one for each sublattice, and in the absence of an electric field one obtains the following self-consistency conditions relating the impurity (or local backbone) Green's functions $G$ and the impurity hybridization function $\Delta$:\cite{Georges1996} $\Delta_{\text{A/B},\sigma}(t,t')=t_\text{hop}^2 G_{\text{B/A},\sigma}(t,t')$, were $G_\text{A/B}$ represents the backbone Green's function on sublattice A/B and $\Delta_\text{A/B}$ the hybridization function of the corresponding impurity problem. We use $t_\text{hop}$ as the unit of energy and $\hbar/t_\text{hop}$ as the unit of time, with $\hbar=1$. 

To study the nonequilibrium properties of photo-doped states, we excite the system with a laser pulse described by the time-dependent electric field $E(t)$ acting along the $t_\text{hop}$ bonds, using the procedure detailed in Ref.~\onlinecite{Werner2017}. Within this ``Bethe lattice plus field" approach, the hopping term acquires a time-dependent Peierls phase, $v(t)=t_\text{hop}e^{\pm i\phi(t)}$, where the sign in the exponent depends on whether the hopping is ``parallel" or ``antiparallel" to the field. Here, $\phi(t)=eaA(t)$, $e$ is the electric charge, $a$ the lattice spacing, and the vector potential $A(t)$ is related to the electric field $E(t)$ by $E(t)=-\partial_t A(t)$. The DMFT self-consistency condition becomes $\Delta_{\text{A/B},\sigma}(t,t')=\frac{1}{2}[v(t)G_{\text{B/A},\sigma}(t,t')v^*(t')+v^*(t)G_{\text{B/A},\sigma}(t,t')v(t')]$. We solve the nonequilibrium impurity models using the non-crossing approximation (NCA).\cite{Keiter1971,Eckstein2010}

\section{Results}
\label{sec:results}

\subsection{Photoemission spectrum}

\subsubsection{Equilibrium system}
\label{sec:equilibrium}

We first analyze the equilibrium spectral functions of the model. Here and in the following, we set $U=5$ and $t_d=2$. Figure~\ref{fig_eq} plots $A_\alpha(\omega)=-\frac{1}{\pi}\text{Im}G_\alpha^R(\omega)$ for different temperatures $T$. The top (bottom) panel shows the results for the backbone site of the $\alpha=$~A (B) sublattice. At the lowest temperature ($T=0.2$), both spectra exhibit a gap size of about 2.5, but the structures of the upper and lower bands differ significantly. The A sublattice spectrum features two sharp peaks at $\omega\approx \pm 2.6$ and smaller peaks at $\omega\approx \pm 6.9$, reminiscent of the spectral function of the isolated dimer,\cite{Petocchi2022} as discussed below. The B sublattice spectrum shows upper and lower Hubbard bands centered at $\omega\approx \tfrac U2=\pm 2.5$, which at low temperature exhibit a dip at the position of the peaks in the A-site spectrum, indicating that the inter-site hopping $t_\text{hop}$ produces a level-splitting at low $T$. As the temperature is increased, the sharp peaks in the A-site spectrum are broadened and the splitting of the Hubbard bands disappears. The A-site spectrum starts to develop a metallic peak inside the gap, while the Mott gap in the B-site spectrum is only slightly filled in. 

The basic features of these spectra can be captured by the atomic solutions ($t_\text{hop}=0$) and the exact diagonalization (ED) result for a three-orbital cluster consisting of the dimer coupled by hopping $t_\text{hop}=1$ to a single site. The results for $T=0.2$ (black lines) and $T=2$ (pink lines) are shown in Fig.~\ref{fig_eq_ED}. We see that already the isolated dimer spectrum reproduces the main peaks of the low-$T$ spectrum and the appearance of the in-gap states on the A sites at high $T$. These features are associated with initial states in the $N=2$ charge sector. The peaks at $\omega\approx \pm 2.7$ are excitations from the $N=2$ ground state $|2,s\rangle=0.61(|\!\uparrow,\downarrow\rangle -|\!\downarrow,\uparrow\rangle)+0.34(|\!\uparrow\downarrow,0\rangle+|0,\uparrow\downarrow\rangle)$ to the ground states (bonding states) of the $N=1$ and $N=3$ sectors, $|1,b\rangle=\frac{1}{\sqrt{2}}(|\sigma,0\rangle+|0,\sigma\rangle)$ and $|3,b\rangle=\frac{1}{\sqrt{2}}({|\sigma,\uparrow\downarrow\rangle}-|\!\!\uparrow\downarrow,\sigma\rangle)$, while the peaks at $\pm 6.7$ are excitations from $|2,s\rangle$ to the excited (anti-bonding) states of these sectors, $|1,a\rangle=\frac{1}{\sqrt{2}}(|\sigma,0\rangle-|0,\sigma\rangle)$ and $|3,a\rangle=\frac{1}{\sqrt{2}}(|\sigma,\uparrow\downarrow\rangle+|\!\uparrow\downarrow,\sigma\rangle)$ (note the minus sign of the hopping term in $H_\text{A,loc}$). The in-gap peaks at $\pm 0.5$ are associated with transitions from the thermally excited half-filled spin-triplet states $|2,t\rangle=\{{|\!\uparrow,\uparrow\rangle}, {|\!\downarrow,\downarrow\rangle},\frac{1}{\sqrt{2}}({|\!\uparrow,\downarrow\rangle}+|\!\downarrow,\uparrow\rangle)\}$ to the ground states of the $N=1,3$ sectors ($|1,b\rangle$, $|3,b\rangle$), while the peaks at $\pm 4.5$ are transitions from $|2,t\rangle$ to excited $N=1,3$ anti-bonding states of the $N=1,3$ sectors ($|1,a\rangle$, $|3,a\rangle$). Because the energy splitting between $|2,s\rangle$ and $|2,t\rangle$ is $\Delta E=2.2$, the latter two features appear only at correspondingly high temperatures. 

On the B site, the atomic spectrum shows little temperature dependence in the range considered, since the gap between the $N=1$ and $N=0,2$ charge sectors is larger than $T$ and there are no excited states in the $N=1$ sector. The ED spectrum however changes substantially once we consider the coupling to the neighboring dimer in the three-orbital calculation. The hopping $t_\text{hop}$ splits the peaks representing the Hubbard bands into bonding and anti-bonding bands, very similar to what is observed in the B-site DMFT spectra in Fig.~\ref{fig_eq}. On the A site, the coupling induced by $t_\text{hop}$ does not eliminate the main peaks (which in this case are associated with bonding states), but results in the appearance of two smaller side-peaks, again in good qualitative agreement with the DMFT spectra. The in-gap states which appear at high temperature in the A-site spectra are not much affected by the inter-site hopping.

\begin{table}[t]
\centering
\begin{tabular}{lllll}
$T$ & $m_\text{A}/h$ \hspace{6mm}\mbox{}& $m_\text{B}/h$ \hspace{6mm}\mbox{}& $d_\text{A}$ \hspace{6mm}\mbox{}& $d_\text{B}$ \\
\hline
$0.1$	& $-1.12$ & $11.3$  & $0.126$ & $0.0166$  \\
$0.2$	& $-0.515$ & $5.19$  & $0.126$ & $0.0167$ \\
$0.5$	& $-0.0766$ & $1.89$   & $0.126$ & $0.0249$ \\
$1$	& $0.283$ & $0.831$  & $0.128$ & $0.0588$ \\
$2$      & $0.274$ & $0.367$  & $0.145$ & $0.119$  \\
\hline
noneq. &&\\
$t=10$ & $0.483$ & $3.70$  & $0.155$ & $0.0742$  \\
$t=20$ & $0.714$ & $3.43$ & $0.142$ & $0.0659$  \\
$t=30$ & $0.817$ & $3.24$  & $0.136$ & $0.0617$ 
\end{tabular}
\caption{Upper rows: Induced magnetization on the backbone orbitals of the A and B sublattices, divided by the uniform applied field $h$ for $h=0.001$, as well as the double occupations in these orbitals at the indicated temperatures $T$. Lowest three rows: magnetization divided by field and double occupations measured in the system with initial $T=0.2$ after an electric field pulse with $E_0=2.5$ and $\Omega=5$.}
\label{tab_moments}
\end{table}

\begin{figure*}[ht]
\begin{center}
\includegraphics[angle=-90, width=0.9\columnwidth]{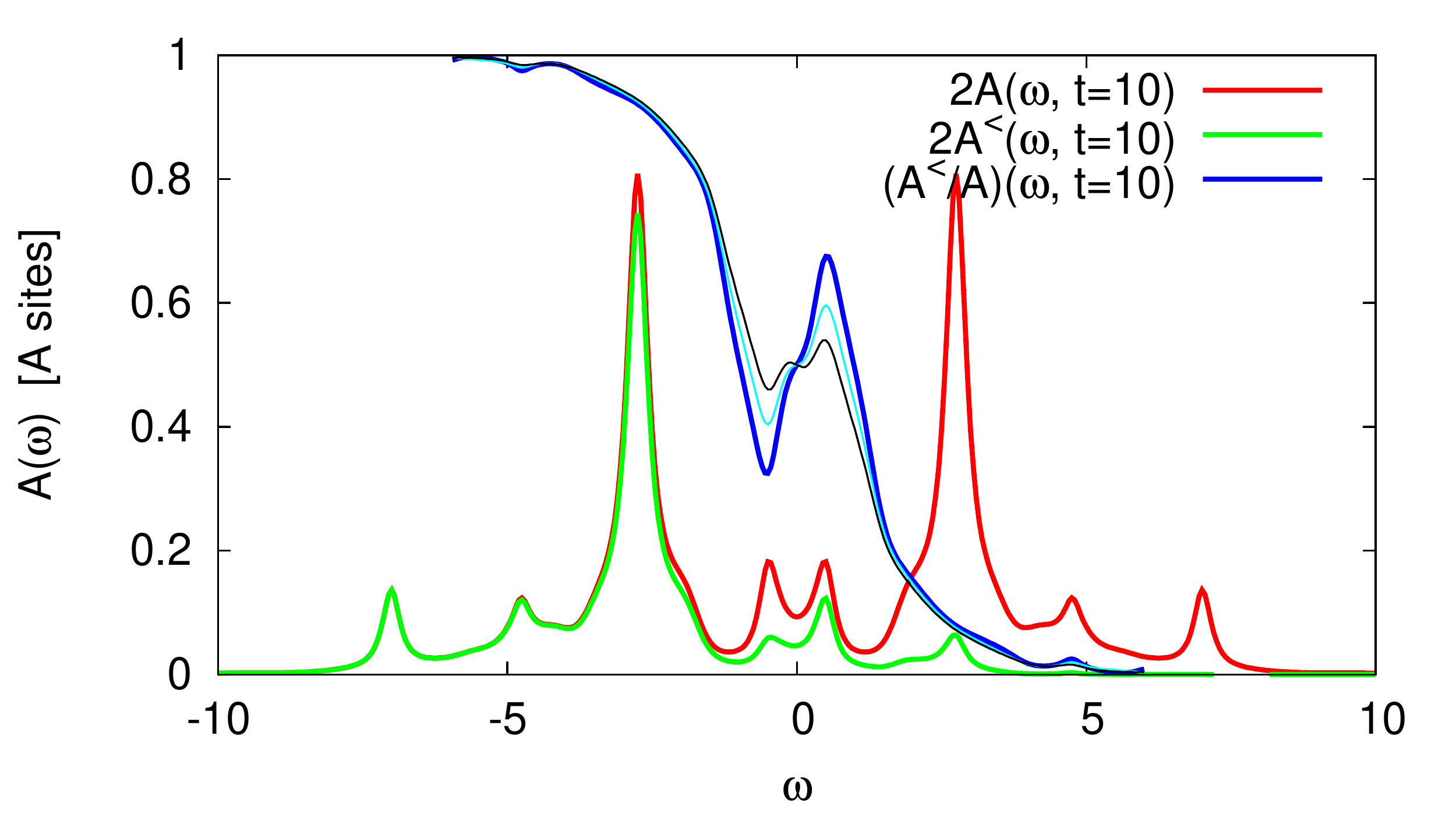} \hspace{10mm}
\includegraphics[angle=-90, width=0.9\columnwidth]{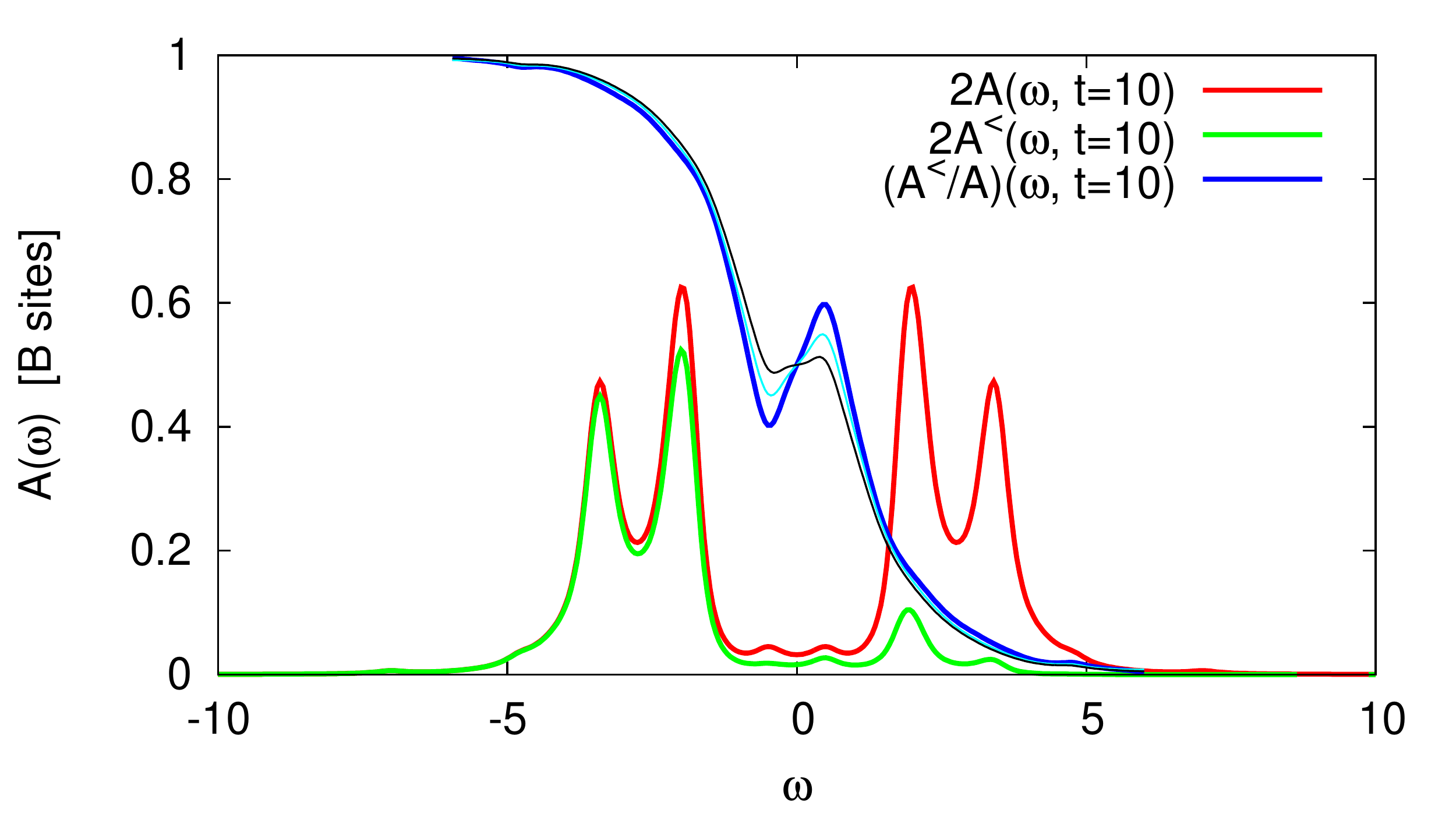} \\ 
\includegraphics[angle=-90, width=0.9\columnwidth]{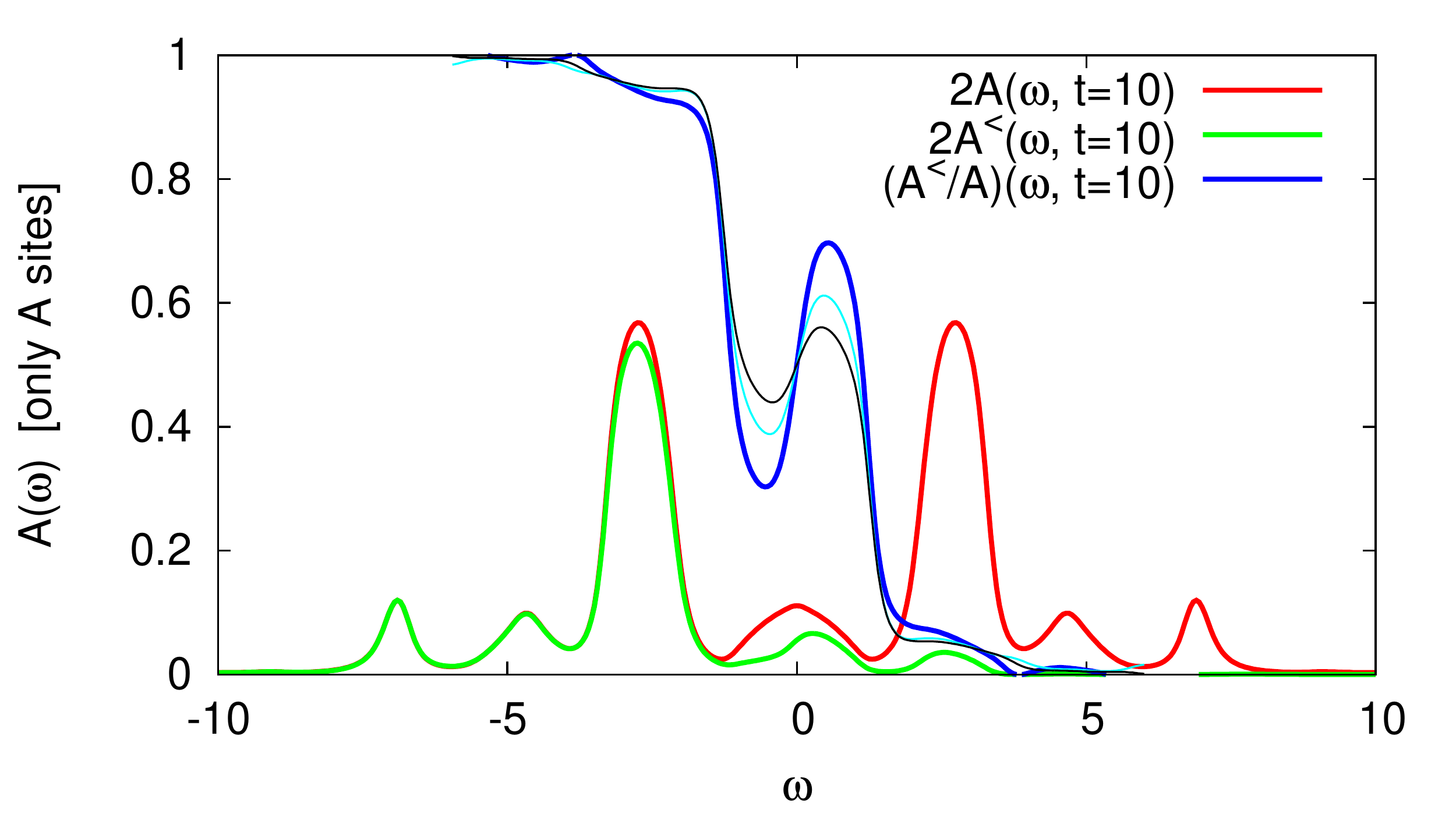} \hspace{10mm}
\includegraphics[angle=-90, width=0.9\columnwidth]{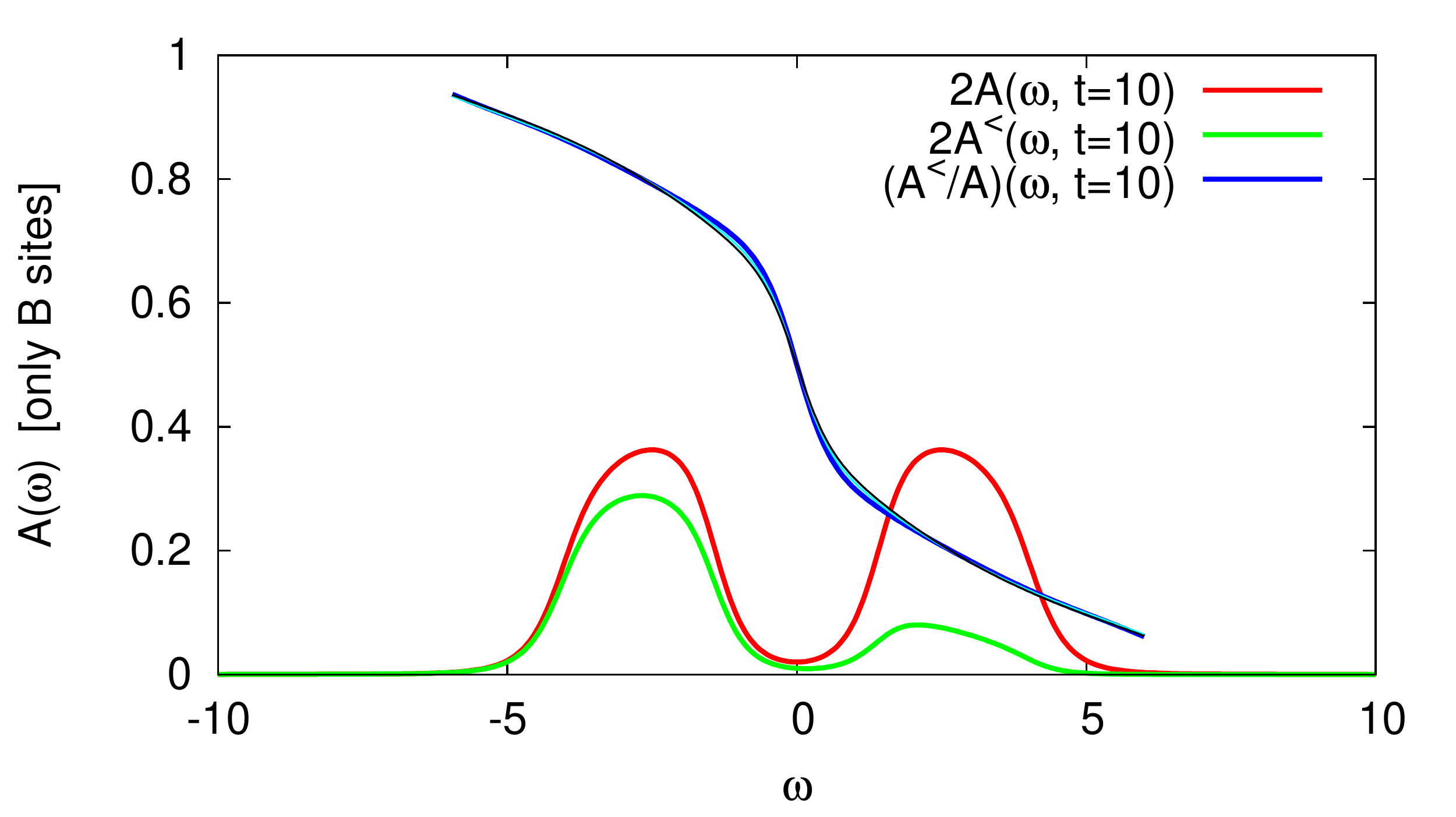} 
\caption{Nonequilibrium spectral functions (red), occupation functions (green), and distribution functions (blue) measured at $t=10$ for the model with $U=5$, $t_d=2$ and an excitation pulse with $E_0=2.5$ and $\Omega=5$ centered at $t_\text{pu}=6$. The thin blue (thin black) lines show the distribution functions measured at $t=14$ ($t=18$). In the top row, we plot the results of our two-sublattice model, with the data for sublattice A in the left panel and the data for sublattice B in the right panel. For comparison, in the bottom row, we also show analogous data for a model consisting only of A-type sites (left) or B-type sites (right). 
}
\label{fig_noneq}
\end{center}
\end{figure*}   

To demonstrate the difference in the magnetic susceptibility on the two sublattices, we apply a small uniform magnetic field $h=0.001$ to all the orbitals and measure the induced magnetization $m_\alpha=n_{\alpha,\uparrow}-n_{\alpha,\downarrow}$ on the backbone site of sublattice $\alpha$. The results are listed in Tab.~\ref{tab_moments} and show the existence of magnetic moments on sublattice B (strong increase in the uniform susceptibility at low $T$), while the moments are suppressed by the dominant singlet state on sublattice A, consistent with the sketch in Fig.~\ref{fig_model} and the ab-initio results for rare earth nickelates in Ref.~\onlinecite{Park2012}.  (The sign change of the magnetization on the A sites at low $T$ is a consequence of the antiferromagnetic exchange with the B sites.) The double occupations $d_\alpha=\langle n_{\alpha,\uparrow} n_{\alpha,\downarrow}\rangle$ listed in the last two columns of the table confirm the picture of A sites in a correlated band-insulator/Kondo-insulator state with a significant probability of double occupations in the backbone orbital, and essentially singly occupied Mott insulating B sites.

\subsubsection{Photo-doped system}

Next, we excite the system at $T=0.2$ with a few-cycle electric field pulse (pump) of the form $E(t)=E_0 f(t-t_\text{pu})\sin(\Omega (t-t_\text{pu}))$, with peak amplitude $E_0=2.5$ and frequency $\Omega=U=5$, centered at $t_\text{pu}=6$ (see Sec.~\ref{sec:model}). The pulse envelope has a Gaussian form, $f(t)=\exp(-t^2/2)$. In the top panels of Fig.~\ref{fig_noneq}, we plot the nonequilibrium spectral functions $A_\alpha(\omega,t)=-\frac{1}{\pi}\int_t^{t_\text{max}}dt' e^{i\omega(t'-t)}G_\alpha^R(t,t')$ and the nonequilibrium occupations $A^<_\alpha(\omega,t)=\frac{1}{2\pi}\int_t^{t_\text{max}}dt' e^{i\omega(t'-t)}G_\alpha^<(t,t')$ for the backbone orbitals on the A (left panel) and B (right panel) sublattice, together with the nonequilibrium distribution function $A^<(\omega,t)/A(\omega,t)$. In the latter case, the thick blue line shows the result for $t=10$, the thin blue line for $t=14$, and the thin black line for $t=18$. At short times after the pulse, the population of the in-gap states appearing in the A- and B-sublattice spectra is inverted, while the population of the higher-energy structures and Hubbard bands corresponds to a relatively well-defined positive effective temperature $T_\text{eff}$, which changes slowly with time and which is very similar on the two sublattices. 

Here, we should mention that in a photo-doped state, we may expect additional A-site spectral features associated with excitations from the $N=1$ and $N=3$ charge sectors to the neighboring $N\pm 1$ sectors. In the case of the dimer (atomic limit for the A sites) one finds that the lowest-energy excitations associated with these transitions are at energy $\omega = \pm 0.5$, which is the same as the peaks associated with excited $N=2$ states. Hence, singly and triply occupied A sites can manifest themselves in the form of population-inverted in-gap states. Similarly, the presence of doublons and holes on the B sites can be associated with a partial population of the upper (lower) Hubbard band with doublons (holes). This is indeed what is seen in the occupations (green curves) and distribution functions (``negative temperature distribution" in the gap region). The evolution of the distribution function shows that the population inversion disappears relatively quickly. This can be related to a relaxation of photo-excited doublon and hole carriers via the in-gap region, which allows for a thermalization of the system towards a global temperature which is much faster than in a large gap Mott insulator (see below).  

For comparison, we also show in the bottom row of Fig.~\ref{fig_noneq} the results for the same pulse applied to a lattice composed only of A-type sites (left panel) or B-type sites (right panel). These calculations correspond to the self-consistency condition $\Delta_{\text{A/B},\sigma}(t,t')=\frac{1}{2}[v(t)G_{\text{A/B},\sigma}(t,t')v^*(t')+v^*(t)G_{\text{A/B},\sigma}(t,t')v(t')]$. The nonequilibrium distribution function for the A-type lattice is similar to that of the original model, while the result for the B-type lattice looks qualitatively different. In particular, the latter distribution function does not show any population inversion in the gap region, where the density of states remains low. We also notice the very broad distribution function in the energy region of the Hubbard bands, which indicates a high effective temperature for the doublons and holons. This is consistent with previous results for the photo-doped paramagnetic single-band Hubbard model, which showed a broad energy distribution and long life-time of the charge carriers, and an absence of photo-induced metallic in-gap peaks.\cite{Eckstein2011,Eckstein2013} 

We can conclude from this comparison that the energy distribution in the original model is primarily controlled by the A-sublattice sites, where the appearance of photo-induced in-gap states enables a relaxation of the photo-carriers into the gap region and a fast recombination. 

\begin{figure}[t]
\begin{center}
\includegraphics[angle=-90, width=0.9\columnwidth]{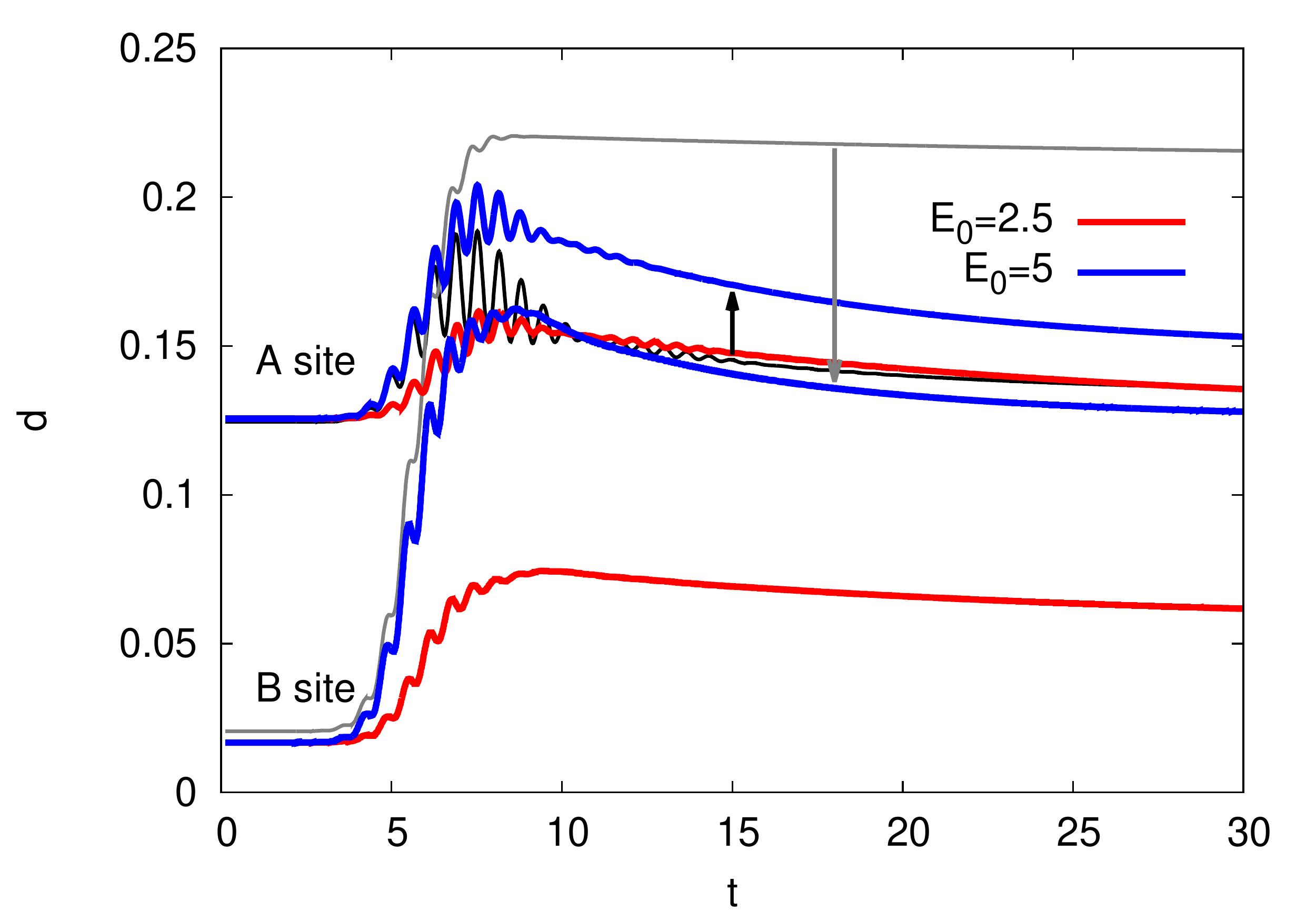} %
\caption{Evolution of the double occupation on the backbone site of the A and B sublattice during and after the pump excitation with amplitude $E_0=2.5$ (red curves) and $E_0=5$ (blue curves). The black (gray) lines show the results for systems with only A-type sites (only B-type sites), and excitation strength $E_0=5$. The corresponding arrows highlight the effect of the coupling between the sublattices on the doublon population after the pulse. 
}
\label{fig_d}
\end{center}
\end{figure}

The time-dependent doublon population on the A and B sites is plotted by the red curves in Fig.~\ref{fig_d} for the pulse with $E_0=2.5$. One can see a significant enhancement during the pulse, especially on the B sites, followed by a decrease in the doublon population on both sublattices. This shows that in the system with coupled A and B sublattices, the doublons and holons on the Mott insulating B sublattice hop to the Kondo-insulating A sublattice, where they can recombine. To show this more clearly, we also plot the results for a stronger pump pulse ($E_0=5$, blue lines), together with the evolution of the double occupation in systems with only A or only B sites (black and gray lines). From the comparison of the two it becomes clear that doublons are transferred from the B sites to the A sites, which results in a reduction of the doublon population on the B sites (compared to the B-site only model) and an enhancement of the doublon population on the A sites. Without the ability to move to (and recombine on) the A sites, the doublon and holon population on the B sites would relax much more slowly, consistent with previous DMFT simulations of comparable single-orbital Mott systems.\cite{Eckstein2011}

The evolution of the magnetization induced by a small constant magnetic field $h=0.001$ is shown for the original model and $E_0=2.5$ in the lowest three rows of Tab.~\ref{tab_moments}. The data indicate a reduction (increase) in the uniform magnetic susceptibility on the B (A) sites due to the photo-carrier injection and a relatively slow approach to a hot thermalized state. Even in the photo-excited state, with about 6-7\% doublons and holons on the Mott insulating sites, our ``bond disproportionated" system exhibits a substantially larger magnetic moment on the B sublattice.

\subsection{RIXS spectrum}

\subsubsection{Equilibrium system}

As a second probe of the equilibrium and photo-doped state we consider resonant inelastic X-ray scattering (RIXS).\cite{Hill1998,Luuk2011} This photon-in, photon-out technique excites electrons from a core level to an empty valence state with the incoming photon with energy $\omega_\text{in}$, and measures the radiation emitted with frequency $\omega_\text{out}$ when the core hole is filled by the decay of a valence electron. Since the RIXS process involves initially unoccupied valence states, it provides complementary information to PES. We will start by calculating the equilibrium RIXS spectra for the A and B sites, and in the following subsections discuss the modifications in the RIXS spectra induced by the previously considered photo-doping pulse. Similar time-resolved RIXS measurements have recently been realized experimentally.\cite{Dean2016,Mitrano2019,Mitrano2020}

To measure the RIXS signal with nonequilibrium DMFT, we add a ``core" orbital with energy shift $\Delta_c$ and interaction strength $U_{cd}$ to each backbone site and simulate the excitation of electrons from the core to the valence orbital via a probe pulse field $E_\text{pr}(t)=E_\text{pr}f_\text{pr}(t-t_\text{pr})\sin(\omega_\text{in}(t-t_\text{pr}))$ with small amplitude $E_\text{pr}$.  Here, $t_\text{pr}$ is the probe time and $f_\text{pr}(t)$ a Gaussian envelope function. The local Hamiltonians on the A and B sites are thus extended to $H_{\alpha,\text{loc}}^\text{RIXS}=H_{\alpha,\text{loc}} - \Delta_c n_{\alpha,c} + U_{cd}n_\alpha(n_{\alpha,c}-2)$ with $c^\dagger_{\sigma}$ representing the creation operator for the core states, $n_{c,\sigma}=c^\dagger_{\sigma}c_{\sigma}$, and $n_c=n_{c,\uparrow} + n_{c,\downarrow}$. The Hamiltonian describing the dipolar excitations between the core and the backbone valence states ($d_{\alpha=\text{A,B}}$ operators) is $H_\text{pr}=E_\text{pr}(t)\sum_\sigma (P^\dagger_{\alpha,\sigma} + P_{\alpha,\sigma})$ with $P_{\alpha\sigma}=c^\dagger_{\alpha,\sigma} d^{\phantom\dagger}_{\alpha,\sigma}$, where we assume that matrix elements are included in the definition of the amplitude $E_\text{pr}$. The DMFT simulation measures the correlation functions $D_{\alpha,\sigma,\sigma'}(t,t')=-i\langle T_\mathcal{C} P^{\phantom\dagger}_{\alpha\sigma}(t)P_{\alpha\sigma'}^\dagger(t')\rangle$, from which the RIXS signal can be calculated.\cite{Eckstein2021} To mimic the short core-hole life-time, we couple a full, noninteracting electron bath with a box-shaped density of states and hopping strength $t_\text{bath}$ to the core orbital (hybridization function $\Delta_\text{bath}(t,t')=t_\text{bath}^2 G_{0,\text{bath}}(t,t')$, with $G_{0,\text{bath}}(t,t')$ the noninteracting Green's function corresponding to the box-shaped density of states). Details of the nonequilibrium DMFT based RIXS formalism  can be found in Ref.~\onlinecite{Eckstein2021}. The present setup is analogous to the one used in a recent time-resolved RIXS study of a two-orbital Hubbard model.\cite{Werner2021} We choose $\Delta_c=27$, $U_{cd}=7$, $E_\text{pr}=0.01$, a Gaussian probe envelope $f_\text{pr}(t)=\exp(-t^2/4)$, a bath density of states covering the energy range $-30 \le \omega \le -10$, and bath coupling strength $t_\text{bath}=2$.

\begin{figure*}[ht]
\begin{center}
\includegraphics[angle=0, width=\columnwidth]{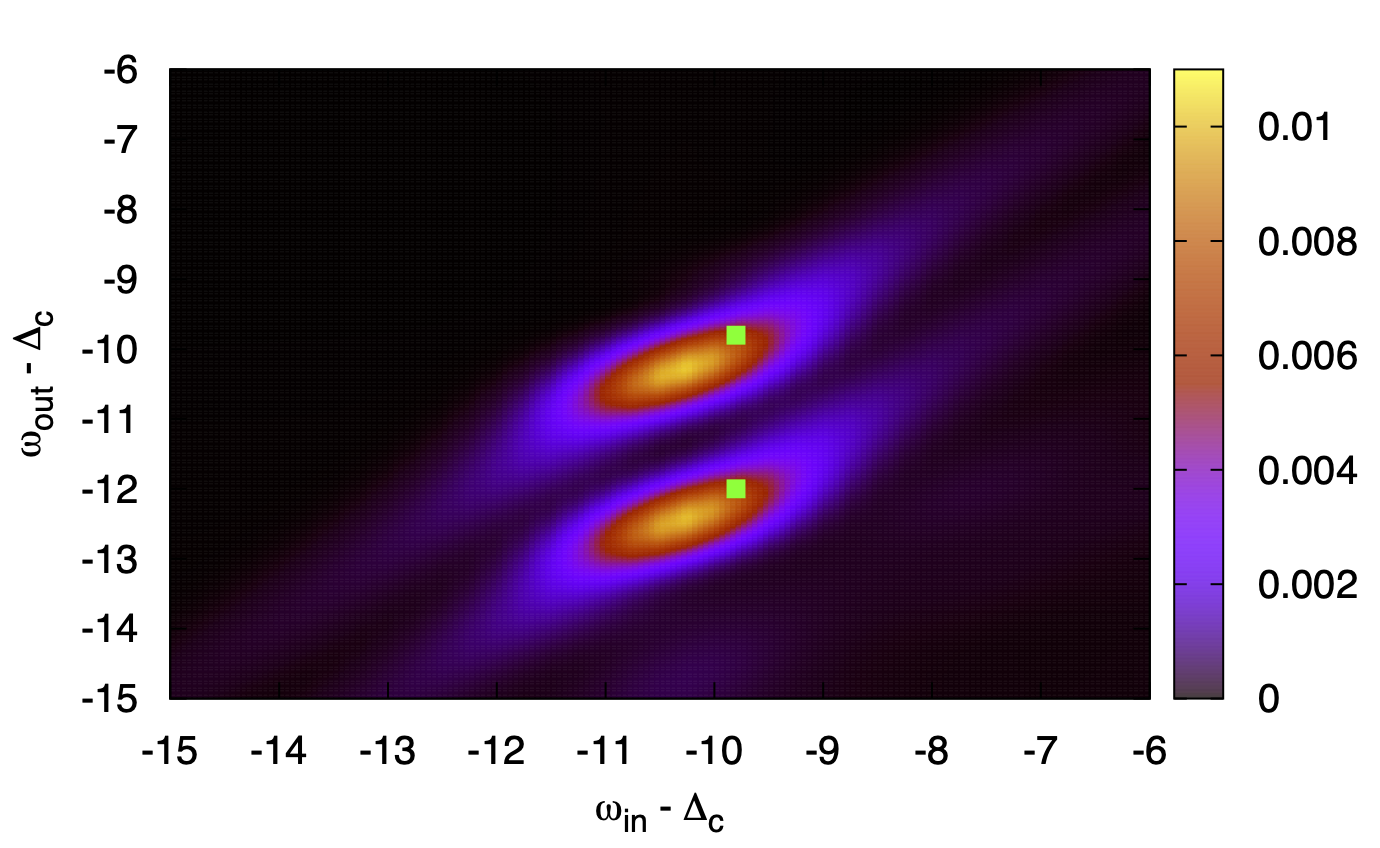} \hspace{0mm} 
\includegraphics[angle=0, width=\columnwidth]{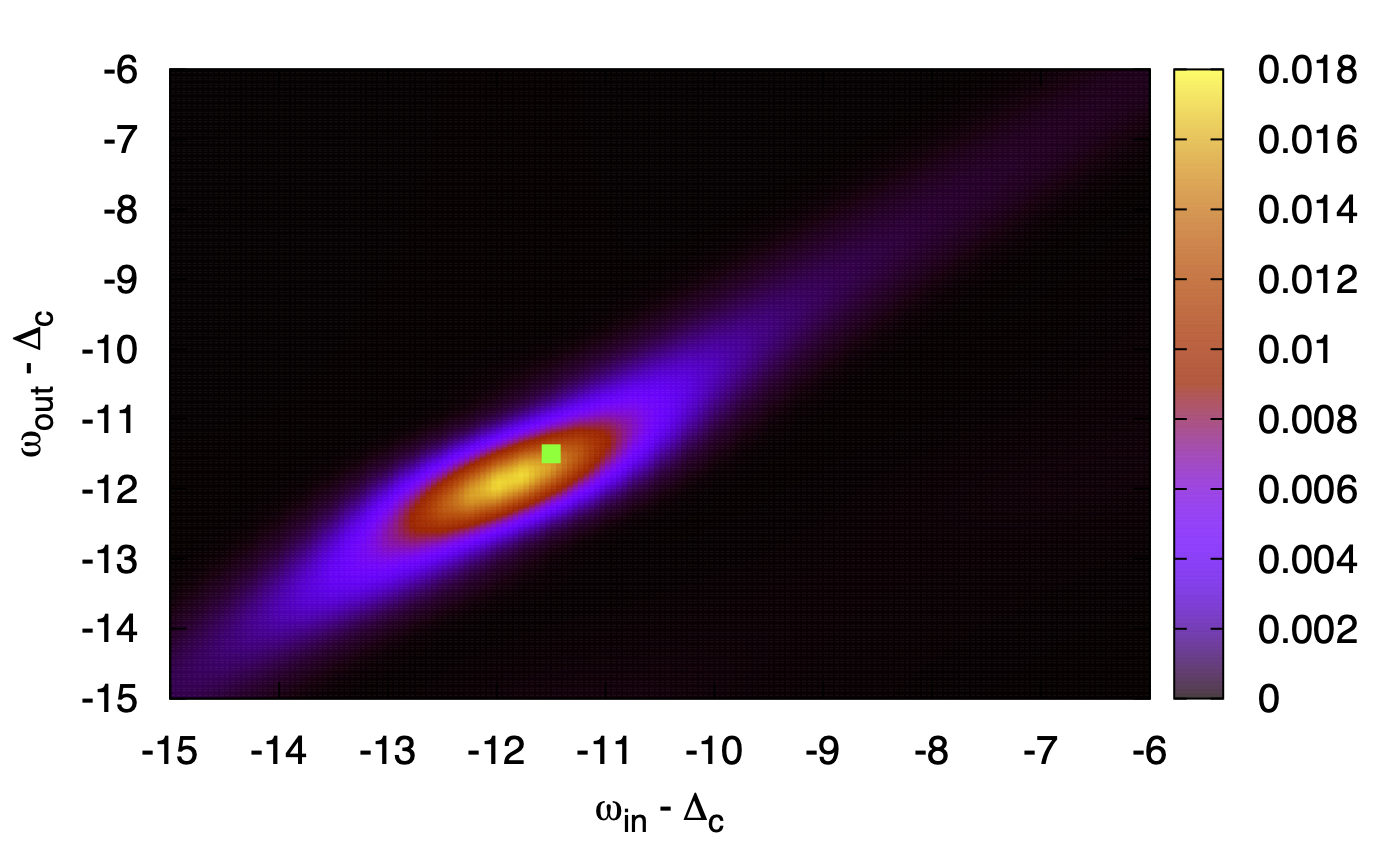} 
\caption{Equilibrium RIXS spectrum (in arbitrary units) measured on the A sublattice (left panel) and B sublattice (right panel).
Green squares show the energies of the RIXS peaks in the ED spectra obtained for $|i\rangle$ the ground state with filling $N=2+n_c=4$ for the dimer (A sites) and $N=1+n_c=3$ for the single orbital (B sites). 
}
\label{fig_rixs}
\end{center}
\end{figure*}   

The equilibrium RIXS signal of the Mott insulating B sublattice, shown in the right panel of Fig.~\ref{fig_rixs}, can be easily interpreted.
 It exhibits mainly the elastic peak associated with excitations of core electrons to the singly occupied backbone orbital. The corresponding intermediate state excitation energy is $\omega_\text{in}-\Delta_c=U/2-2U_{cd}=-11.5$. The de-excitation process primarily converts this doubly occupied backbone orbital back into a singly occupied one (elastic scattering) and thus the emitted radiation is expected to peak near the same energy $\omega_\text{out}-\Delta_c=-11.5$.  At elevated temperatures, one may additionally expect a very weak signal associated with transitions to an empty backbone orbital, at energy $\omega_\text{in}-\Delta_c=\omega_\text{out}-\Delta_c=-U/2-U_{cd}=-9.5$, while the doubly occupied B site orbitals are not detectable. The measured RIXS spectrum is consistent with these atomic-limit considerations, although the peak signal is elongated along the elastic line $\omega_{\text{in}}=\omega_{\text{out}}$ due to the short core-hole life-time. The slight tilt with respect to the diagonal can be related to relaxation processes of the intermediate states due to lattice effects and the interplay between the core decay and the finite probe pulse duration. Since the $U_{cd}$ interaction leads to a strongly bound exciton, the doublon produced by the core excitation cannot easily hop to other sites, and hence we do not see any pronounced loss features associated with the creation of excitations on neighboring sites. 
 
On the A sites, the situation is a bit more complicated, since there is a loss peak at $\omega_\text{in}-\omega_\text{out}\approx 2.2$, which can be associated with a singlet-triplet excitation. To interpret the DMFT result shown in the left panel of Fig.~\ref{fig_rixs}, it is useful to calculate the RIXS spectrum for the atomic system using the Kramers-Heisenberg formula\cite{Kramers1925,Matsubara2005} 
\begin{align}
I^\text{RIXS}_\alpha(\omega_\text{in},\omega_\text{out}) \propto & \sum_f \Bigg|
\sum_{m,\sigma} \frac{\langle f | c^\dagger_{\alpha,\sigma}d_{\alpha,\sigma} | m\rangle\langle m |d^\dagger_{\alpha,\sigma}c_{\alpha,\sigma} |i\rangle}{\omega_\text{in}+E_i-E_m+i\Gamma}
\Bigg|^2\nonumber\\
&\times\delta(\omega_\text{in}+E_i-\omega_\text{out}-E_f),
\label{eq_kramers}
\end{align}
where $|i\rangle$ is the initial state (ground state) with energy $E_i$, and $|m\rangle$, $|f\rangle$ are the intermediate and final states with energies $E_m$, $E_f$, respectively. 
$\Gamma$ is a small broadening parameter related to the core bath in our real-time formulation.\cite{Eckstein2021} The green squares in Fig.~\ref{fig_rixs} show the dominant peaks of the equilibrium RIXS spectrum, obtained in the atomic limit with the half-filled ground state as the initial state $|i\rangle$. In the case of the B sublattice, where $|i\rangle=|\sigma\rangle_d |\!\!\uparrow\downarrow\rangle_c$, we obtain the expected peak at $\omega_\text{in}-\Delta_c=\omega_\text{out}-\Delta_c=-11.5$, while on the A sublattice, where $|i\rangle=|2,s\rangle_d |\!\!\uparrow\downarrow\rangle_c$, one finds two peaks. (We use the same notation for the valence states as introduced in Sec.~\ref{sec:equilibrium}.) The elastic feature is at $\omega_\text{in}-\Delta_c=\omega_\text{out}-\Delta_c=-9.8$, corresponding to $|f\rangle=|i\rangle $ (half-filled ground state of the dimer plus filled core level), and the loss feature is at $\omega_\text{in}-\Delta_c=-9.8$, $\omega_\text{out}-\Delta_c=-12$, corresponding to $|f\rangle=|2,t\rangle_d|\!\uparrow\downarrow\rangle_c$ (three degenerate half-filled spin-triplet states of the dimer plus filled core level). The peaks in the DMFT RIXS signal are shifted with respect to the green squares because of lattice effects and the coupling to the core bath. The broadening along the diagonal is again controlled by the short life-time of the core hole.   

\subsubsection{Photo-doped system}

\begin{figure*}[t]
\begin{center}
\includegraphics[angle=0, width=\columnwidth]{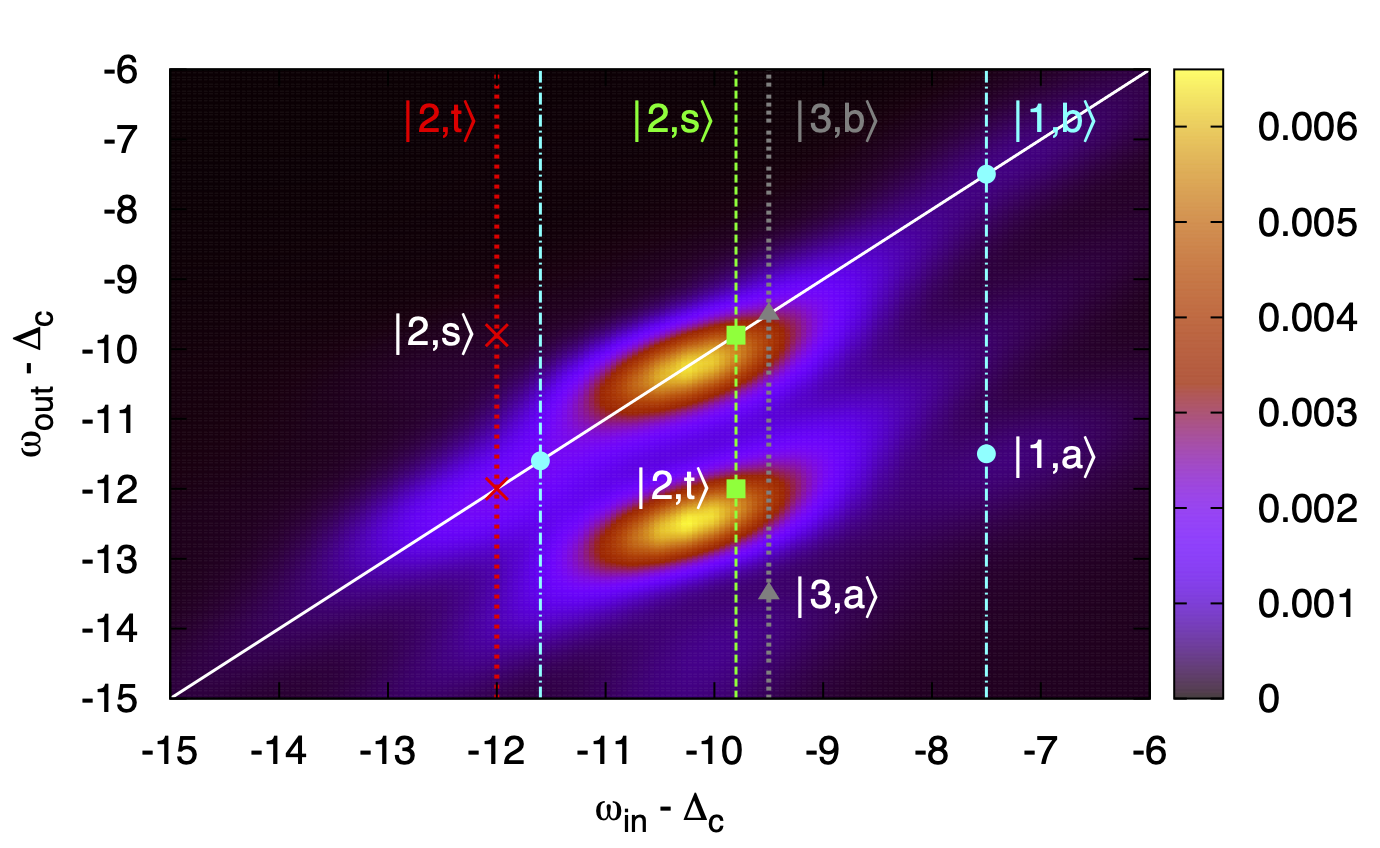} \hspace{0mm}
\includegraphics[angle=0, width=\columnwidth]{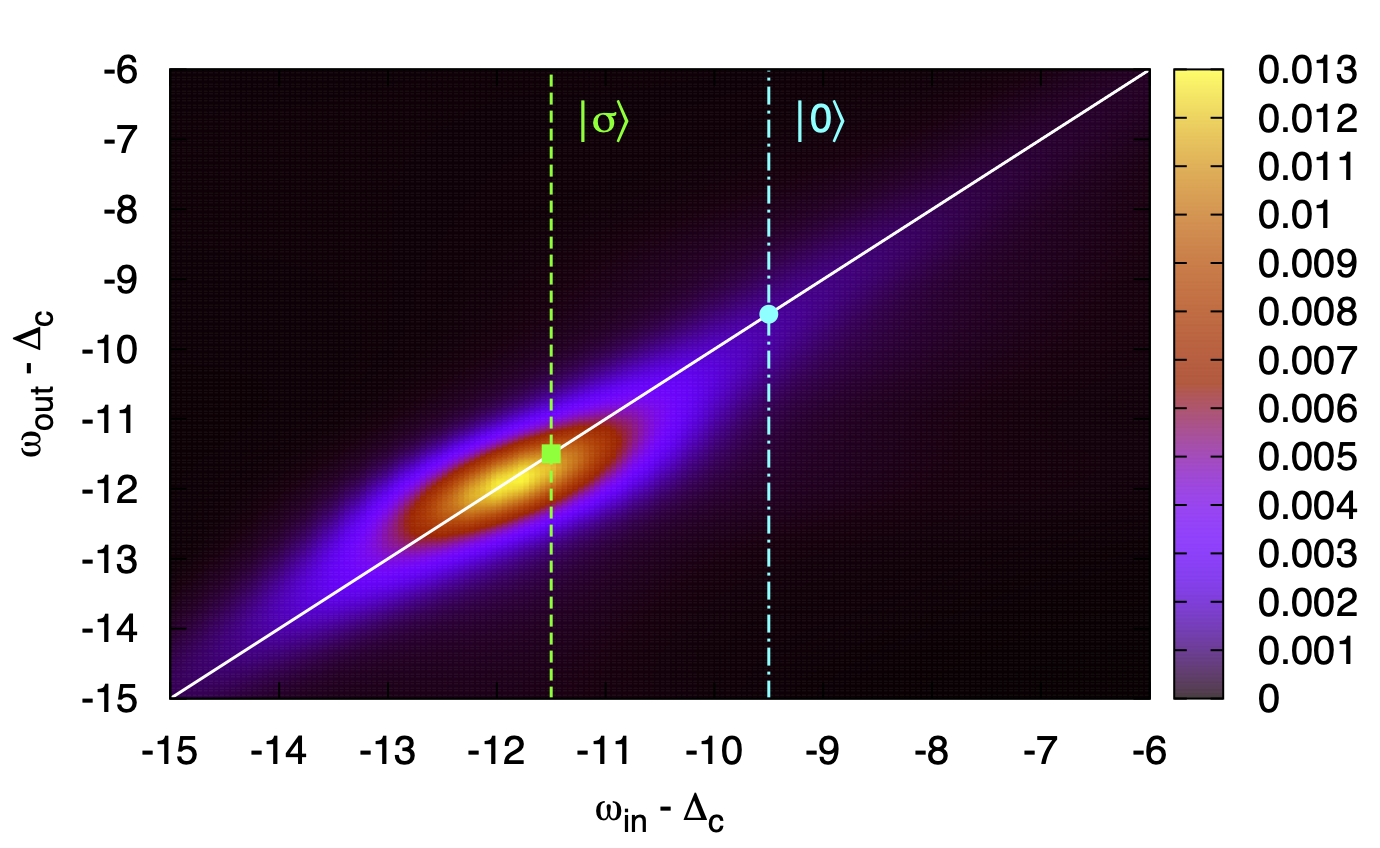}\\
\includegraphics[angle=0, width=\columnwidth]{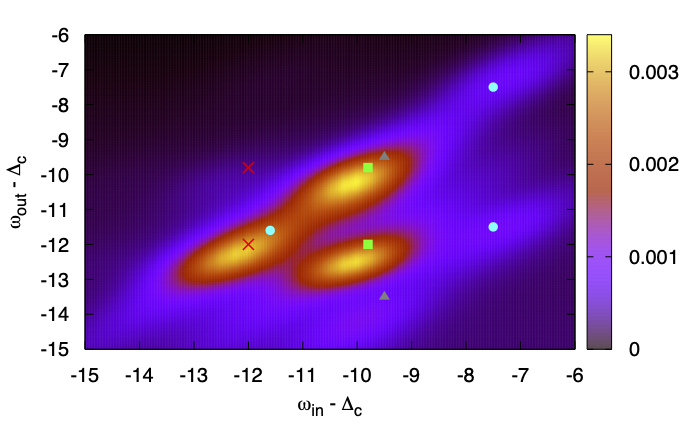} \hspace{0mm}
\includegraphics[angle=0, width=\columnwidth]{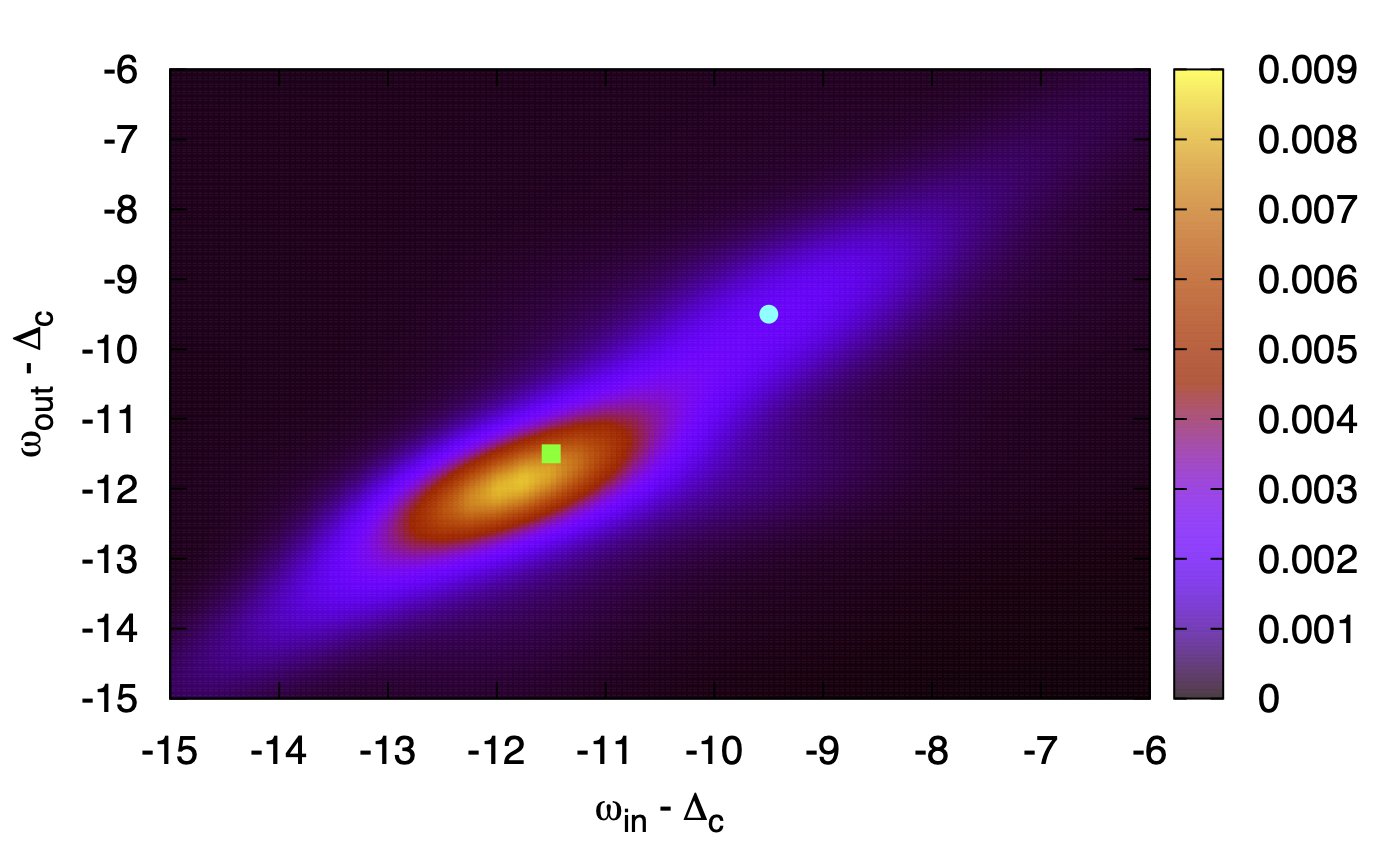} 
\caption{Nonequilibrium RIXS spectrum measured at $t_\text{pr}=10$ on the A sites (left panels) and B sites (right panels) after the pump pulse centered at $t_\text{pu}=6$. The top panels show the results for pump field amplitude $E_0=2.5$ and the bottom panels for $E_0=5$.
Green squares, light-blue circles and gray triangles show the energies of the RIXS peaks in the ED spectra obtained for $|i\rangle$ the ground state  with filling $N$, $N-1$ and $N+1$, respectively. Here, $N=2+n_c=4$ for the dimer (A sites) and $N=1+n_c=3$ for the single orbital (B sites). The red crosses in the A site panels are obtained by choosing $|i\rangle$ as an excited triplet state in the $N$ sector. In the top panels, we indicate the valence states $|i\rangle_d$ associated with $|i\rangle$ in the colors of the corresponding vertical lines. The valence states $|f\rangle_d$ of the final configurations $|f\rangle$ are indicated for the gain and loss features by the white labels. 
}
\label{fig_rixs_pumped}
\end{center}
\end{figure*}

The top panels of Fig.~\ref{fig_rixs_pumped} show the spectra after photo-excitation with pump pulse amplitude $E_0=2.5$, and $t_\text{pu}=6$, measured at probe time $t_\text{pr}=10$. Apart from an overall reduction in the intensity of the RIXS signal, we notice the appearance of additional peaks in the RIXS spectrum, and a reduced tilt of the main features (less relaxation of the electrons within the bands before the filling of the core hole). On the B sites, the signal gets enhanced around $\omega_\text{in}-\Delta_c=\omega_\text{out}-\Delta_c=-9.5$ (see light blue dot in the top right panel), which is the expected energy for the holon peak. This signal is weak, but in line with the increase in the long-lived doublon and holon population by only about $4\%$ (see red lines in Fig.~\ref{fig_d} and population of the upper Hubbard band in Fig.~\ref{fig_noneq}). This small increase is due to the fact that the absorption in a system with A and B sublattices is weaker than in the B-only system, and because doublons and holons can hop to the A sites where they recombine.

The A sites, which are not only excited by the pump pulse, but also absorb the photo-carriers hopping in from the B sublattice, show a more significant change in the RIXS spectrum. For pump pulse amplitude $E_0=2.5$ (top left panel), there appear additional peaks, which can be associated with photo-doped or exited initial states $|i\rangle$. 
In the figure, we indicate the peak positions (obtained for the isolated dimer-plus-core systems from Eq.~\eqref{eq_kramers}), which can be associated with the ground states of the different charge sectors. Choosing $|i\rangle=|2,s\rangle_d|\!\uparrow\downarrow\rangle_c$ as the ground state of the half-filled dimer ($N=2+n_c=4$) yields the green squares, choosing $|i\rangle=|1,b\rangle_d|\!\!\uparrow\downarrow\rangle_c$ as the ground state of the hole-doped dimer ($N=1+n_c=3$) produces the light-blue circles, while the ground state $|i\rangle=|3,b\rangle_d|\!\uparrow\downarrow\rangle_c$ of the electron-doped dimer ($N=3+n_c=5$) yields the gray triangles. The corresponding peaks in the DMFT RIXS spectra can be clearly identified, although they are shifted in energy by lattice effects. Note that in the case of the hole-doped dimer, two different intermediate states in the RIXS process $|i\rangle \rightarrow \{|m_1\rangle, |m_2\rangle\}\rightarrow |f\rangle$ result in two different $\omega_\text{in}$-values for the poles. The final states for the poles corresponding to gain or loss features are indicated in the top panels of the figure by the white labels.  

The additional weak features which are evident in the DMFT spectra must be related to excited states $|i\rangle$ in the different charge sectors. Particularly relevant are the triplet states $|2,t\rangle_d|\!\uparrow\downarrow\rangle_c$ of the half-filled ($N=2+2=4$) sector, which yield two peaks: a dominant elastic peak at ($\omega_\text{in}-\Delta_c$, $\omega_\text{out}-\Delta_c$)=($-12$,$-12$) and a much weaker gain feature at  ($\omega_\text{in}-\Delta_c$, $\omega_\text{out}-\Delta_c$)=($-12$,$-9.8$) associated with a return to the ground state (``anti-Stokes" process). These energies, which are indicated by red crosses, suggest that the strong signal appearing around ($\omega_\text{in}-\Delta_c$, $\omega_\text{out}-\Delta_c$)=($-12$,$-12$) is primarily due to the population of spin-triplet states in the photo-excited system.   

The bottom panels of Fig.~\ref{fig_rixs_pumped} show analogous nonequilibrium RIXS spectra measured at $t_\text{pr}=10$ after a stronger pump pulse with $E_0=5$ (see the blue lines in Fig.~\ref{fig_d} for the doublon production). The left panel shows the A-site spectra and the right panel the B-site spectra. We notice the same photo-induced features as in the system excited by the $E_0=2.5$ pulse, but the additional peaks have now more weight, relative to the main peaks, and are thus more clearly visible in the intensity plots. In the case of the B-site spectrum, one starts to recognize a weak loss feature near ($\omega_\text{in}-\Delta_c$, $\omega_\text{out}-\Delta_c$)=($-9.5$,$-11.5$), which appears to be strongly tilted (fluorescent line). This feature corresponds to the excitation of the core electron to an empty B site, the hopping in of an electron from a neighboring A site, and the filling of the core hole from the resulting doubly occupied state. If the electron is hopping in from a singlon site, then the atomic configurations on the two sites are equivalent before and after the RIXS process, so that the energy loss corresponds to a transfer of kinetic energy to the system. If the electron hops in from a neighboring doublon, there would be an additional gain of energy $U$, which makes this process very unlikely. Hence, even in the presence of a high density of doublons and holons, the system with large $U_{cd}$ will not exhibit a strong feature associated with doublon-holon recombination processes. 

The spectra for the A-only system (not shown) are a bit less blurred than for the two-sublattice system, consistent with the lower density of photo-induced doublons in Fig.~\ref{fig_d} (no doublons and holons hopping in from B sites), while in the B-only system, the holon feature is enhanced relative to the main singlon peak, because the production of a large density of doublons and holons depletes the singlon population, and the doublons and holons cannot hop to A sites. Apart from these differences associated with the transfer of doublons and holons between sublattices, the RIXS spectrum of the A-only and B-only systems look similar to the RIXS spectra for the respective sites in the two-sublattice system. 

\subsubsection{Time evolution of the photo-doped system}

\begin{figure*}[ht]
\begin{center}
\includegraphics[angle=0, width=\columnwidth]{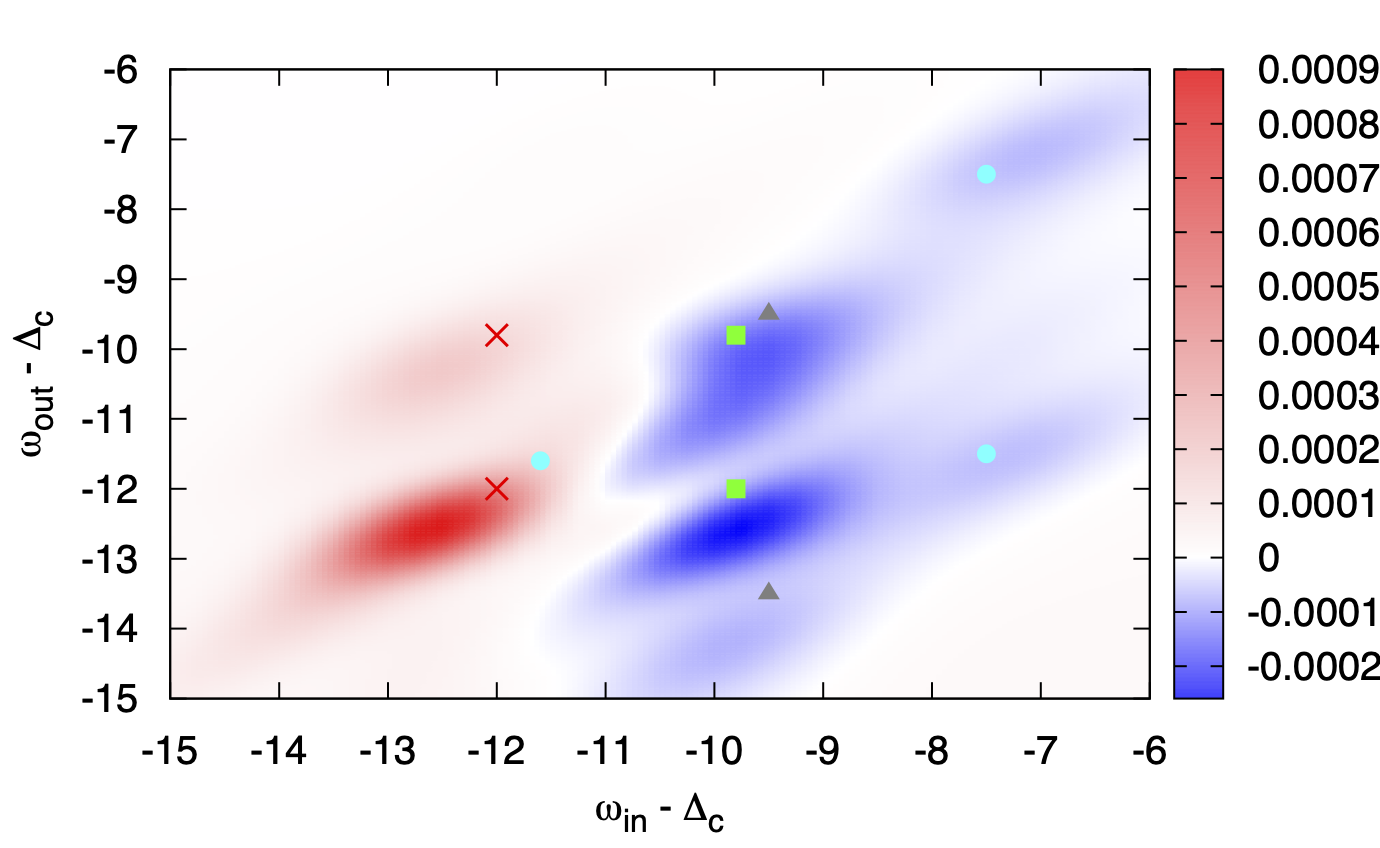} \hspace{0mm}
\includegraphics[angle=0, width=\columnwidth]{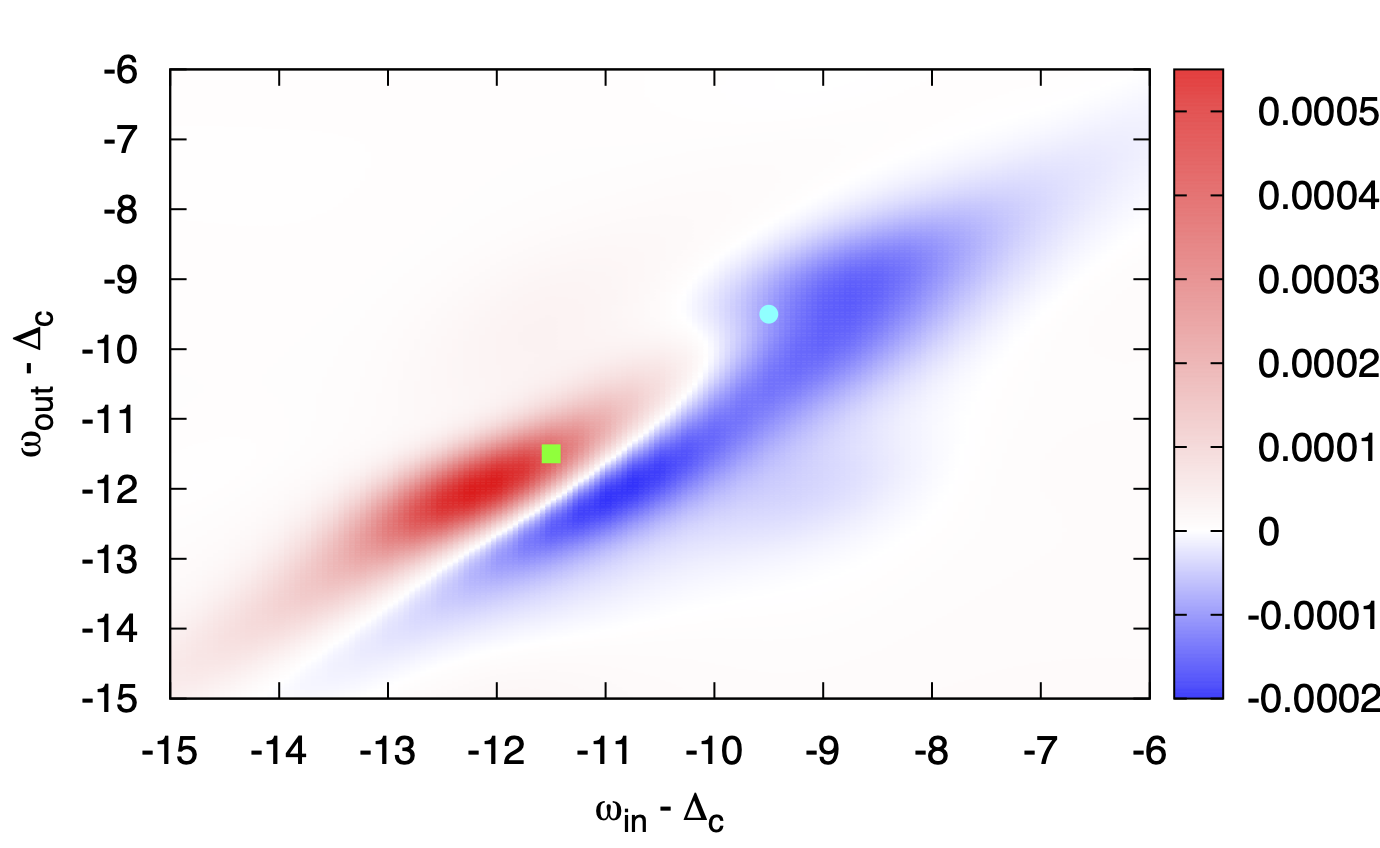} 
\caption{
Difference between the nonequilibrium RIXS spectra for $E_0=5$ measured at $t_\text{pr}=14$ and $t_\text{pr}=10$, $I^\text{RIXS}_{\alpha=1}(t_\text{pr}=14)-I^\text{RIXS}_{\alpha=1}(t_\text{pr}=10)$. The left panel is for the A sites and the right panel for the B sites. Here, the red color corresponds to an increasing RIXS signal and blue to a decreasing RIXS signal. Green squares, light-blue circles and gray triangles show the energies of the RIXS peaks in the ED spectra obtained for $|i\rangle$ the ground state  with filling $N$, $N-1$ and $N+1$, respectively. Here, $N=2+n_c=4$ for the dimer (A sites) and $N=1+n_c=3$ for the single orbital (B sites). The red crosses in the A site panel are obtained by choosing $|i\rangle$ as an excited triplet state in the $N$ sector.  
}
\label{fig_rixs_diff}
\end{center}
\end{figure*}   

As was demonstrated in Ref.~\onlinecite{Werner2021}, the time evolution of the nonequilibrium RIXS signal reveals information on the evolution of the local state populations. Figure~\ref{fig_rixs_diff} shows the difference between the RIXS spectra for pump field amplitude $E_0=5$ measured at probe times  $t_\text{pr}=14$ and $t_\text{pr}=10$. On the B sites, the evolution of the RIXS spectrum indicates an increase in the weight of the peak at $\omega_\text{in}-\Delta_c=\omega_\text{out}-\Delta_c=-11.5$, associated with singly occupied backbone orbitals (green square), and a decrease in the weight of the peak at $\omega_\text{in}-\Delta_c=\omega_\text{out}-\Delta_c=-9.8$ which comes from empty orbitals (light blue dot). This is consistent with the hopping of holes and doublons from the B sites to the A sites, as is evident also in Fig.~\ref{fig_d} by comparing the evolution of the doublons in the B-site-only system (gray line) to that in the two-sublattice system (blue line). In contrast, the evolution of the A site spectrum reveals an increase in the peak associated with the triplet states $|2,t\rangle_d |\!\uparrow\downarrow\rangle_c$ of the half-filled ($N=2+2=4$) sectors, both for the elastic and gain features (see red crosses), while the signals associated with the doubly occupied ground state (green squares) are losing weight. This is consistent with an excited state population which is still growing on the A sites at $t=14$, due to the absorption of doublons and holes from the neighboring B sites. We note that there is no evidence of a growing population of triply occupied dimers (gray triangles), which indicates that the corresponding increase is more than compensated by the decrease in the half-filled populations with overlapping peak energies, or that the triplon-singlon recombination is very fast.

\section{Conclusions}
\label{sec:conclusions}

Motivated by the physics of rare earth nickelates and 1$T$-Ta$S_2$, where band-insulating or Kondo-insulating subsystems coexist with Mott insulating subsystems, we have studied the PES and RIXS signals of a model whose equilibrium state is characterized by a bonding/antibonding gap and spin-singlet state on the A sublattice and a Mott insulating state with local moment on the B sublattice. We have shown that the coupling of the two different types of insulating solutions by the hopping $t_\text{hop}$ leads to a splitting of the Hubbard bands (B site) and the appearance of side-peaks of the bonding/antibonding peaks (A sites) in the PES signal at low temperature. At high temperature, or in photo-doped states, in-gap peaks associated with spin triplet states of the dimer appear in the A site spectrum, which becomes metallic, while the B site spectrum remains essentially gapped. 

If an electric field pulse with frequency $\Omega\approx U$ is applied to this system, doublon-holon pairs are generated on the Mott insulating B sublattice. Within a time of a few inverse hoppings, a substantial fraction of these charge carriers moves to the A sublattice, where they populate the in-gap states and recombine. Together with the photo-carriers produced on the A sublattice, this inter-sublattice transfer of photo-carriers contributes to a transient population inversion in the low-energy region. This results in a nonequilibrium distribution function which is similar on both sublattices, but primarily controlled by the dynamics of the excited state population on the A sublattice. 

Our analysis shows that the qualitatively different nature of the insulators on the two sublattices can be revealed by very hot electronic temperatures or nonthermal electron populations. From a site-averaged local PES signal it is however difficult to extract the population dynamics on the individual sublattices or the flow of charge carriers between the sublattices. The RIXS spectrum can provide additional information, in particular if the relevant peaks from the A and B sites are well-separated in the $\omega_\text{in}$-$\omega_\text{out}$ plane, and if theoretical modeling allows to assign the peaks to the different sublattices.  For example, the equilibrium RIXS signal on the B sublattice exhibits a dominant elastic feature associated with excitations of core electrons to singly occupied states, while the equilibrium spectrum of the A sublattice exhibits a shifted elastic peak associated with half-filled initial and final states of the dimer, and a loss feature associated with the creation of excited triplet states in the half-filled sector. In the photo-doped system, a characteristic feature originating from the presence of holons appears on the B sites, and this peak is well-separated from the equilibrium and photo-doped signals associated with the A sublattice. This feature allows to track the holon (and by symmetry doublon) population on the Mott insulating sublattice, which is controlled by the hopping of these charge carriers to the neighboring A sublattice. 

On the A sites, the influx of charge carriers enhances the nonequilibrium features of the RIXS spectrum, in particular the signals associated with excited spin-triplet states of the dimer. The corresponding elastic peak and gain feature are separated in the $\omega_\text{in}$-$\omega_\text{out}$ plane from the peaks associated with the B sublattice and thus can be used to extract information on the time evolution of the spin-triplet states on the A sublattice. From an analysis of the time-dependent changes in these peak weights it may be possible to reconstruct the dynamics of the local state populations\cite{Werner2021} and from there the transfer of charge carriers between the sublattices.    

The information on the elastic line, including the evolution of excited state populations, can in principle also be extracted from simpler X-ray absorption spectroscopy measurements.\cite{Werner2022} The gain and loss features in the RIXS spectrum additionally reveal the energies associated with excitation and de-excitation processes within the valence manifold. 

While our model calculations provide a proof-of-principles, the investigation of realistic materials, such as rare earth nickelates or 1$T$-TaS$_2$, of course requires a proper ab-initio derived model for the valence states and a more accurate treatment of the core levels and transition matrix elements. Such calculations are computationally more expensive, but in principle doable, at least with approximate solvers such as the NCA used in this work. They may be helpful for guiding or interpreting future time-resolved PES and RIXS studies of correlated insulators which feature a nontrivial interplay between band-insulating/Kondo-insulating and Mott insulating behavior, or shed light on materials where the nature of the insulating state is not yet fully understood.

\acknowledgments

We thank T. Schmitt for helpful discussions. P.~W. and F. P. acknowledge support from ERC Consolidator Grant No. 724103 and SNSF Grant No. 200021-196966. M. E. is supported by ERC Starting Grant No.~716648. The calculations were run on the beo05 cluster at the University of Fribourg using a code based on the NESSi library.\cite{Nessi}

\bibliographystyle{eplbib}

\begin{thebibliography}{99}

\bibitem{Medarde1997} M. Medarde, J. Phys: Cond. Matt. {\bf 9}, 1679 (1997).

\bibitem{Alonso1999} J. A. Alonso, J. L. Garcia-Munoz, M. T. Fernandez-Diaz, M. A. G. Aranda, M. J. Martinez-Lope, and M. T. Casais
Phys. Rev. Lett. {\bf 82}, 3871 (1999).

\bibitem{Mizokawa2000} T. Mizokawa, D. I. Khomskii, and G. A. Sawatzky, Phys. Rev. B {\bf 61}, 11263 (2000).

\bibitem{Park2012} H. Park, A. J. Millis, and C. A. Marianetti, Phys. Rev. Lett. {\bf 109}, 156402 (2012).

\bibitem{Lau2013} B. Lau and A. J. Millis, Phys. Rev. Lett. {\bf 110}, 126404 (2013).

\bibitem{Johnston2014} S. Johnston, A. Mukherjee, I. Elfimov, M. Berciu, and G. A. Sawatzky, Phys. Rev. Lett. {\bf 112}, 106404 (2014).

\bibitem{Mazin2007} I. I. Mazin, D. I. Khomskii, R. Lengsdorf, J. A. Alonso, W. G. Marshall, R. M. Ibberson, A. Podlesnyak, M. J. Martinez-Lope,
and M. M. Abd-Elmeguid, Phys. Rev. Lett. {\bf 98}, 176406 (2007).

\bibitem{Subedi2015} A. Subedi, O. E. Peil and A. Georges, Phys. Rev. B {\bf 91}, 075128 (2015).

\bibitem{Georges1996} A. Georges, G. Kotliar, W. Krauth, and M. J. Rozenberg, Rev. Mod. Phys. {\bf 68}, 130 (1996).

\bibitem{Aoki2014} H. Aoki, N. Tsuji, M. Eckstein, M. Kollar, T. Oka, and P. Werner, Rev. Mod. Phys. {\bf 86}, 779 (2014).

\bibitem{footnote_L} We choose the same interaction $U$ in the ligand and backbone orbitals for simplicity. The main role of the interaction on the L orbitals is to favor singly occupied configurations.

\bibitem{Ritschel2018} T. Ritschel, H. Berger, and J. Geck, Phys. Rev. B {\bf 98}, 195134 (2018).

\bibitem{Lee2019} S.-H. Lee, J. S. Goh, and D. Cho, Phys. Rev. Lett. {\bf 122}, 106404 (2019).

\bibitem{Petocchi2022} F. Petocchi, C. W. Nicholson, B. Salzmann, D. Pasquier, O. V. Yazyev, C. Monney, and P. Werner, Phys. Rev. Lett. {\bf 129}, 016402 (2022). 

\bibitem{Metzner1989} W. Metzner and D. Vollhardt, Phys. Rev. Lett. {\bf 62}, 324 (1989).

\bibitem{Werner2017} P. Werner, H. Strand, S. Hoshino, and M. Eckstein, Phys. Rev. B {\bf 95}, 195405 (2017).

\bibitem{Keiter1971} H. Keiter and J. C. Kimball, Int. J. Magn. {\bf 1}, 233 (1971).

\bibitem{Eckstein2010} M. Eckstein and P. Werner, Phys. Rev. B {\bf 82}, 115115 (2010).

\bibitem{Eckstein2011} M. Eckstein and P. Werner, Phys. Rev. B {\bf 84}, 035122 (2011).

\bibitem{Eckstein2013} M. Eckstein and P. Werner, Phys. Rev. Lett. {\bf 110}, 126401 (2013).

\bibitem{Hill1998} J. Hill, C.-C. Kao, W. Caliebe, M. Matsubara, A. Kotani, J. Peng, and P. Greene, Phys. Rev. Lett. {\bf 80}, 4967 (1998).

\bibitem{Luuk2011} L. Ament , M. van Veenendaal, T. Devereaux, J. Hill, and J. van den Brink, Rev. Mod. Phys. {\bf 83}, 705 (2011).

\bibitem{Dean2016} 
M. Dean {\it et al.}, Nature Mat. {\bf 15}, 601 (2016). 

\bibitem{Mitrano2019} M. Mitrano {\it et al.}, Science Advances {\bf 5}, eaax3346 (2019). 

\bibitem{Mitrano2020}
M. Mitrano and Y. Wang, Commun. Phys. {\bf 3}, 184 (2020). 

\bibitem{Eckstein2021} M. Eckstein and P. Werner, Phys. Rev. B {\bf 103}, 115136 (2021).

\bibitem{Werner2021} P. Werner, S. Johnston and M. Eckstein, Europhys. Lett. {\bf 133}, 57005 (2021).

\bibitem{Kramers1925} H. A. Kramers and W. Heisenberg, Z. Phys. {\bf 31}, 681 (1925).

\bibitem{Matsubara2005} M. Matsubara, T. Uozumi, A. Kotani, and J. C. Parlebas, J. Phys. Soc. Jpn. {\bf 74}, 2052 (2005).

\bibitem{Werner2022} P. Werner, D. Golez and M. Eckstein, arXiv:2204.06762 (2022).

\bibitem{Nessi} M. Sch\"uler, D. Golez, Y. Murakami, N. Bittner, A. Herrmann, H. U.R. Strand, P. Werner, and M. Eckstein, Computer Physics Communications {\bf 257}, 107484 (2020).

\end{thebibliography}

\end{document}